\documentclass[
aip,
jcp,
floatfix,
 amsmath,amssymb,
 reprint,%
]{revtex4-1}

\usepackage{graphicx}
\usepackage{dcolumn}
\usepackage{bm}
\usepackage{soul}
\usepackage[utf8]{inputenc}
\usepackage[T1]{fontenc}
\usepackage{mathptmx}
\usepackage{etoolbox}
\usepackage{mathastext}

\usepackage{pgfplots}
\usepackage{tikz}
\usepackage{subfigure}
\usetikzlibrary{shapes,arrows,chains,fit,calc,shapes,matrix,positioning,patterns,decorations.pathreplacing,calligraphy}
\usepgfplotslibrary{fillbetween}
\pgfplotsset{compat=newest}
\usepgfplotslibrary{groupplots}
\usetikzlibrary{external}
\usepackage{comment}
\usepackage{multirow}
\usepackage{subfigure}
\usepackage{xr}
\usepackage{hyperref}

\makeatletter
\def\@email#1#2{%
 \endgroup
 \patchcmd{\titleblock@produce}
  {\frontmatter@RRAPformat}
  {\frontmatter@RRAPformat{\produce@RRAP{*#1\href{mailto:#2}{#2}}}\frontmatter@RRAPformat}
  {}{}
}%
\makeatother

\begin{document}

\preprint{AIP/123-QED}

\title{The Role of the Pre-exponential Factor on Temperature Programmed Desorption spectra: a Computational Study of Frozen Species on Interstellar Icy Grain Mantles}
\author{S. Pantaleone}
\thanks{S. Pantaleone and L. Tinacci contributed equally to this work}
\affiliation{Università degli Studi di Torino, Dipartimento di Chimica, via P. Giuria 7, 10125 Torino, Italy}
\author{L. Tinacci}
\thanks{S. Pantaleone and L. Tinacci contributed equally to this work}
\affiliation{Università degli Studi di Torino, Dipartimento di Chimica, via P. Giuria 7, 10125 Torino, Italy}
\affiliation{Université Grenoble Alpes, CNRS, IPAG, 38000 Grenoble, France}
\author{V. Bariosco}
\affiliation{Università degli Studi di Torino, Dipartimento di Chimica, via P. Giuria 7, 10125 Torino, Italy}
\affiliation{Departament de Qu\'{i}mica, Universitat Aut\`{o}noma de Barcelona, Bellaterra, 08193, Catalonia, Spain}
\author{A. Rimola}
\affiliation{Departament de Qu\'{i}mica, Universitat Aut\`{o}noma de Barcelona, Bellaterra, 08193, Catalonia, Spain}
\affiliation{Accademia delle Scienze di Torino, Via Maria Vittoria 3, 10123 Torino, Italia}
\author{C. Ceccarelli}
\affiliation{Université Grenoble Alpes, CNRS, IPAG, 38000 Grenoble, France}
\author{P. Ugliengo}
\thanks{Corresponding author: Piero Ugliengo -  piero.ugliengo@unito.it}
\email{piero.ugliengo@unito.it}
\affiliation{Università degli Studi di Torino, Dipartimento di Chimica, via P. Giuria 7, 10125 Torino, Italy}

\date{\today}

\begin{abstract}
Temperature programmed desorption (TPD) is a well-known technique to study gas-surface processes, and it is characterized by two main quantities: the adsorbate binding energy and the pre-exponential factor. While the former has been well addressed in recent years by both experimental and computational methods, the latter remains somewhat ill-defined and different schemes have been proposed in the literature for its evaluation. In the astrochemistry context, binding energies and pre-exponential factors are key parameters which enter microkinetic models for studying the evolution over time of the chemical species in the Universe. In this paper we studied, by computer simulations, the effect of different pre-exponential factor models using as test cases water, ammonia and methanol adsorbed on amorphous and crystalline ices: specifically, the most widely used by the astrochemical community (Herbst-Hasegawa), those provided by Tait and Campbell, and an extension of the Tait formulation including the calculation of the vibrational partition function. We suggest the methods proposed by Tait and Campbell that provide TPD temperature peaks within 30 K to each other while avoiding demanding quantum mechanical calculations as they are based on tabulated data. Finally, when the explicit inclusion of the vibrational partition function is needed, we propose a cost-effective strategy to include all the thermal contributions in the partition functions without the need for performing a full vibrational calculation of the whole system.
\end{abstract}

\maketitle

\section{Introduction}\label{sec:intro}
 
The desorption rate is the fundamental observable needed to describe the microscopic mechanisms that account for the desorption of whatever species adsorbed on a surface.\cite{Cuppen_2024_review_ices} 
In other words, it is the main actor that rules whether a species remains stuck on a surface or goes into the gas phase.
The desorption rate ($k_{des}$) is mathematically described as an Arrhenius process:
\begin{equation}
        \mathrm{k_{des}} = \nu \exp \bigl(-E_a / (RT)\bigr)\; ,
\end{equation}
where the pre-exponential factor $\nu$ is related to the entropy of the desorption mechanism and $\mathrm{E_a}$ is the activation energy, usually approximated as the binding energy (BE). Therefore, BE is the energy needed to detach the adsorbate from the surface. Rigorously, it should include the zero point-point energy (ZPE) and thermal correction to be converted in the heat of desorption.

The determination of the pre-exponential factor (elsewhere also called prefactor) in the chemical desorption mechanism is a long standing issue \cite{fair1980low,campbell2012entropies, Ligterink_desorption_2023} because of the high level of accuracy required in both experimental and computational studies. 
In fact, from the computational point of view, despite several methods provide a very accurate estimate of the energetic term,\cite{piccini2015accurate,sauer2019ab} the prefactor still remains a critical quantity.\cite{fair1980low,sprowl2016hindered}

The main problem connected to the estimation of the prefactor is the difficulty in describing the vibrational modes of the species involved before and after desorption, especially those deriving from the free translations and rotations of the adsorbate that become intermolecular vibrational modes once the gas phase molecule is stuck on the surface. 
In the last decades, different approaches have been proposed to retrieve this fundamental piece of information: the immobile particle approximation\cite{fair1980low,tait2005n,tait2006n} (where the vibrational frequencies are not considered at all), the harmonic oscillator (in which the nuclear motion is limited to small amplitude within a parabolic potential), and the free and hindered translators/rotators.\cite{sprowl2016hindered}
Many approaches were proposed to go beyond the above-mentioned approximations,\cite{jorgensen2017adsorbate,bajpai2018benchmark,waitt2021adsorbate} the one with the most general applicability taking into account the full anharmonic treatment of all vibrational modes, solving the 1D nuclei Schr\"odinger equation for each normal mode and any combination of them.\cite{piccini2013quantum,piccini2014effect,rybicki2022rigid}
This treatment is important, especially for large amplitude motions and low frequencies, and for high-frequency stretchings involving light atoms, in particular H.
A way to overcome the limitations imposed by the static calculations is to take advantage of Molecular Dynamics (MD) simulations\cite{fichthorn2002thermal} which, through the Fourier transform of the velocity autocorrelation function, provide the anharmonic frequency spectrum of the system, \cite{vaf} albeit considering the nuclei as classical particles. Finally, a recent empirical method developed by Campbell \& Sellers \cite{campbell2012entropies, campbell2013kinetic} was introduced to determine the prefactor based on the gas-phase entropy of the adsorbed species. This approach was calibrated using the adsorption of various alkanes on MgO(100) smoke, Pt(111), and Graphite(001). Their findings indicated that the adsorbate retains approximately two-thirds of the entropy of the gas-phase molecule, suggesting that the procedure could be applicable to other surfaces and molecular species.\cite{campbell2012entropies} This has been confirmed by subsequent papers which demonstrated the validity of such method to an extensive set of molecules and materials, not only when adsorbed on flat surfaces, but also on microporous structures like zeolites.\cite{dauenhauer2018universal,rzepa2020computational,budi2018calculation}

However, all of these techniques have been applied to van der Waals systems, \textit{ i.e.} adsorbates bound to metallic, metal oxide, zeolite and Metallic-Organic-Framework substrates ("hard" surfaces) throughout dispersive forces, which does not change the structure and properties of the ad-molecule and the substrate upon adsorption: the assumption of a rigid surface works well, the cohesive energy of the surface being orders of magnitude higher than the interactions that govern the adsorption process.
This may not hold when the interactions between the adsorbate and the substrate are of the same order of magnitude of the cohesive energy of the substrate. Relevant cases are the adsorption through H-bond interactions of adsorbates with the surface of amorphous ice of interstellar relevance where the coupling of the vibrational modes between the ad-molecule and the adsorbate affects the vibrational partition function. It is part of the purpose of this work to assess the relevance of the above hypothesis.

From an experimental point of view desorption rate and related parameters, \textit{i.e.} binding energy and pre-exponential factor, are derived from Temperature Programmed Desorption (TPD) experiments which, applied to "soft" surfaces, may cause the restructuring and/or the co-desorption of the surface molecules, in addition to the adsorbate, because the interactions between the adsorbate and the surface are of the same nature and magnitude of the interactions responsbile of the cohesive energy of the surface itself. Therefore, being very difficult to selectively desorb the adsorbate only, the measures on these kinds of systems often overestimates the desorption rate.\cite{minissale2022thermal}

In astrochemistry this is particularly important because in certain astronomical environments like molecular clouds, the majority of the surfaces available to trap/release gas-phase molecules are the icy mantles covering the rocky core of the grains. The accurate estimate of desorption rates of interstellar molecules is fundamental to obtain reliable predictions of the evolution of such species through the microkinetic models.

In this work, we will focus on the interactions of ammonia, water, and methanol as test cases with icy water surfaces with the aim of simulating adsorption at the icy mantles of interstellar grains. The objective is not to establish new benchmarks with experiments but to compare the prefactor formula commonly employed by the astrochemical community, proposed by Hasegawa  \& Herbst,\cite{Hasegawa1992} the immobile particle approximation proposed by Tait et al.,\cite{tait2005n,tait2006n} its extension throughout the inclusion of the vibrational partition function, thus considering the eventually non-negligible vibrational couplings of adsorbate/adsorbent and the empirical formula proposed by Campbell \& Sellers.\cite{campbell2012entropies} 
The challenge is to apply the different approaches to the adsorption of species at "soft" ice surfaces, with respect to "hard" metallic surfaces where these theories has been widely applied, and discuss the differences. 
We will give some suggestions to optimize the cost-accuracy of the pre-exponential factor, depending on the instruments and resources available. Moreover, we will illustrate a way to fit the desorption rates in order to derive temperature dependent prefactors and BEs in a way that can be easily implemented in microkinetic simulation programs.

\section{Desorption rate constant}\label{sec:rate}

\subsection{Formalism}
 
The desorption process can be formulated with the Polanyi-Wigner equation:\cite{king}
\begin{equation}\label{eq:WP}
    -\frac{dN}{dt} = N^i \underbrace{\nu \exp \bigl(-E_a / (R~ T)\bigr)}_{k_{des}(T)} \; ,
\end{equation}
where $N$ is the number of adsorbed species on the surface, \textit{i} is the order of the process, $\nu$ is the pre-exponential factor (also called prefactor), $E_a$ the activation energy, R the ideal gas constant, T the temperature and t the time. 
Typically, physisorption of small molecules is a non-activated process, so that the activation energy is assumed to be equal to the BE ($E_a \approx BE$) of the adsorbed species. This approximation allows to avoid the expensive search of the transition state structure.

Usually, in the theoretical and experimental astrochemical community, the prefactor is estimated via the following equation proposed by Hasegawa \& Herbst:\cite{Hasegawa1992}

\begin{equation}\label{eq:HH}
    \nu_{HH} = \sqrt{\frac{2~ E_a}{\pi^2 ~m_M ~A}} \; ,
\end{equation}

where A is the surface area per adsorbed molecule (normally assumed to be 10$^{-19}$ m$^2$, and this value is kept constant along this work) and m$_M$ is the mass of the adsorbed species. 
This equation usually leads to a $\nu_{HH}$ range of $10^{12}-10^{13} s^{-1}$.\cite{Hasegawa1992} As an alternative, in some cases the prefactor can be inferred by experiments, running different temperature ramps in the same TPD experiment.\cite{luna2015experimental}
However, the majority of BEs and $\nu$ present in literature are obtained at a given temperature peak and, consequently, they are constant, \textit{i.e.}  without any dependency on the temperature.\cite{minissale2022thermal}

Using the Transition State Theory (TST) we can model the desorption rate as a unimolecular process $C \rightarrow M + S$ (C is the adsorbed complex, M and S are respectively the isolated adsorbed molecule and surface), where the activation barrier is assumed to be equal to the BE (corrected for the ZPE and thermal contributions, \textit{i.e.} $BH(T)$, see Appendix \ref{app:be} for details). 
The associated rate can be written as:
\begin{equation}\label{eq:k_des}
    k_{des}(T) = \nu_{TST}(T) \exp \Biggl(-\frac{BH(T)}{R ~T}\Biggr) \; ,
\end{equation}
where the prefactor $\nu_{TST}(T)$ is expressed as:
\begin{equation}
    \nu_{TST}(T) = \frac{k_B ~T}{h}\frac{^\ddagger q}{q} \kappa \; ,
\end{equation}
and where $^\ddagger q$ and $q$ are the partition functions of the transition state (TS) and the complex C, and $\kappa$ the transmission coefficient that, for an irreversible process such as the desorption, can be assumed to be equal to 1.\cite{voter1985dynamical}.
In the approximation that $E_a \approx BE$ and, accordingly, the transition state is approximated to the product (\textit{i.e.} the non- interacting ad-molecule and surface, M + S), the above equation can be expanded by factorizing the different terms of the internal partition function for each involved species, \textit{i.e.} the adsorbed molecule M, the surface S and the complex C, leading to:
\begin{equation}\label{eq:tst_pre_exponential}
    \nu_{TST}^{vib}(T) = \frac{k_B~ T}{h}{^\ddagger q_{trans}^{2D}(M)} {^\ddagger q_{rot}(M)} \underbrace{\frac{{^\ddagger q_{vib}(M)} {^\ddagger q_{vib}(S)}}{q_{vib}(C)}}_{q_{vib}^{TST}} \; .
\end{equation}
Here we assume $q_{elec} = 1$, which represents a good approximation for closed shell species in the ground state.
The $k_B T / h$ term represents the translational degree of freedom that leads the system to the TS (gas phase), \textit{i.e.} the translational motion of the desorbing molecule normal to the surface. 
In other words, it represents the third missing dimension of the 2D translational partition function ($^\ddagger q_{trans}^{2D}(M)$). 
$k_B T / h$ is the key actor in the estimation of the prefactor and, at a temperature around $100K$, relevant for the astrochemistry conditions, it has a value of $\approx 2\times 10^{12}$ s$^{-1}$.
The translational partition function ($^\ddagger q_{trans}^{2D}(M)$) is:
\begin{equation}
    ^\ddagger q_{trans}^{2D}(M) = \Biggl(\frac{2 ~\pi ~m_M ~k_B ~T}{h^2} \Biggr) A \; ,
\end{equation}
in which A has the same meaning as in Eq.~\ref{eq:HH} (\textit{i.e.} 10$^{-19}$ m$^2$, assuming a monolayer coverage).
The rotational partition function ($^\ddagger q_{rot}(M)$) can be expressed within the rigid rotor approximation:
\begin{equation}
    ^\ddagger q_{rot}(M) = \frac{\sqrt{\pi}}{\sigma ~h^3} \bigl(8 ~\pi^2 ~k_B ~T \bigr)^{\frac{3}{2}} \sqrt{I_x ~I_y ~I_z} \; ,
\end{equation}
where $I_x$, $I_y$, $I_z$ are the principal moment of inertia and $\sigma$ is the symmetry factor that classically identifies the indistinguishable rotational configurations of a specific molecule. In the specific case $\sigma = 1, 2, 3$ for methanol (C$_s$ symmetry group, \textit{i.e.} no rotation axis), water (C$_{2v}$, \textit{i.e.} symmetry axis of order 2) and, ammonia (C$_{3v}$, \textit{i.e.} symmetry axis of order 3), respectively.
Finally, the $q_{vib}$ in the harmonic approximation is obtained as:
\begin{equation}\label{eq:q_vib}
    q_{vib} = \Pi_i^{3N-6} \frac{1}{1 - \exp \Bigl( \frac{h ~\nu_i}{k_B ~T} \Bigr)} \; .
\end{equation}

Eq.~\ref{eq:tst_pre_exponential} is usually approximated in the limit of a completely immobile particle in the adsorbed state \cite{fair1980low,tait2005n,tait2006n,minissale2022thermal} assuming $q_{vib}^{TST}$ equal to one, thus leading to:
\begin{equation}\label{eq:apprx_pre_exponential}
    \nu^{Tait}_{TST}(T) = \frac{k_B~ T}{h}
    \underbrace{\Biggl(\frac{2 ~\pi ~m_M ~k_B ~T}{h^2} \Biggr) A}_{^\ddagger q_{trans}^{2D}(M)}
     \underbrace{\frac{\sqrt{\pi}}{\sigma ~h^3} \bigl(8~ \pi^2 ~k_B ~T \bigr)^{\frac{3}{2}} \sqrt{I_x~ I_y ~I_z}}_{^\ddagger q_{rot}(M)} \; .
\end{equation}

This approximation relies on the assumption that the vibrational modes of the surface and the ad-molecule do not change upon adsorption/desorption and, accordingly, the vibrational partition function ratio of Eq. \ref{eq:tst_pre_exponential} ($q_{vib}^{TST}$) is equal to 1. This can be safely applied to rigid surfaces, whose cohesive energy is orders of magnitude greater than the interaction energy of the ad-molecule with the surface, \textit{i.e.} there is no coupling between the vibrations of the ad-molecule and the surface.
However, for "soft" surfaces (\textit{i.e.} where the molecules are connected by intermolecular weak forces, like the H-bond pattern in the ice), it is possible that the coupling between the vibration of the adsorbate and the phonons of the surface play a non negligible role to the $q_{vib}^{TST}$ term, in particular when considering the adsorption of a molecule occurring through H-bond interactions.
The major contributor that leads to different results with respect to the approximated form are the 6 hindered rotations and translations of the adsorbed molecule (6HRT) converted to internal vibrations upon adsorption. 
Eq. \ref{eq:apprx_pre_exponential} does not depend on the final structure and, therefore, it is not associated to a specific site and BE, giving a constant number for each species. 
Indeed, as reported by Ref. \onlinecite{fair1980low}, the prefactor calculated by Eq.~\ref{eq:apprx_pre_exponential} should be considered as an upper limit. 
Within the TST model the vibrational partition function ratio, $q_{vib}^{TST}$, can be expanded as:
\begin{equation}\label{eq:q_vib_ratio}
\begin{split}
    \frac{{^\ddagger q_{vib}(M)}{^\ddagger q_{vib}(S)}}{q_{vib}(C)} &\approx \underbrace{\frac{{^\ddagger q_{vib}(M)}}{q_{vib}^{intra}(M//C)}}_{q_{vib}^{TST}(M)} \underbrace{\frac{{^\ddagger q_{vib}(S)}}{q_{vib}(S//C)}}_{q_{vib}^{TST}(S)} \frac{1}{q_{vib}^{6HRT}} \; ,
\end{split}
\end{equation}
where $q_{vib}^{6HRT}$ is the vibrational partition function only considering the six hindered rotations and translations of the adsorbed species on the surface, $q_{vib}^{intra}(M//C)$ is the partition function for the 3N-6 intramolecular vibrations of the adsorbed species on the surface and $q_{vib}(S//C)$ the partition function of the surface vibrational modes excluding the contributes from the adsorbed species.

The last model for the prefactor has been proposed by \citet{campbell2012entropies}, some years after the works published by Tait \textit{et al.} \cite{tait2005n,tait2006n}, with a new way to estimate the prefactor based on the entropy of the adsorbed species (S$^\circ_{ad}$) which, in turn, depends on the entropy of the molecule in the gas phase (S$^\circ_{gas}$):
\begin{equation}
\label{eq:prefactor_campebel}
\begin{split}
    \nu_{Campbell} = &\frac{k_B T}{h} \exp \Bigg\{ 
    \frac{0.30 S^\circ_{gas}}{R} + 3.3 -\\
    &-\frac{1}{3} \Bigg[ 18.6 + \ln \bigg( \bigg(\frac{m_A}{m_{Ar}}\bigg)^{\frac{3}{2}} 
    \bigg(\frac{T}{298\,\text{K}}\bigg)^{\frac{5}{2}} \bigg) \Bigg] 
    \Bigg\}
\end{split}
\end{equation}
where m$_A$ is the molar mass of the gas, m$_{Ar}$ is the molar mass of the Argon gas, and S$^\circ_{Ar,298K}$ is the entropy of Ar gas at 1 bar and 298K (=18.6R). See Appendix \ref{app:campbell} for details. Campbell's Eq. \ref{eq:prefactor_campebel} was derived by assuming a surface standard-state coverage of 0.01 monolayer.\cite{campbell2016equilibrium}

\subsection{TPD Spectra and Temperature Peak}\label{subsec:TPD}

TPD spectra are computed by using Eq.~\ref{eq:WP}.
For this reason, the time is expressed as a function of the temperature through the experimental surface heating rate ($\beta$ = dT/dt) obtaining the first-order equation at the monolayer coverage regime (\textit{i.e.} \textit{i}=1):
\begin{equation}\label{eq:tpd}
    -\frac{dN}{dT} = \frac{N}{\beta} \underbrace{\nu(T) \exp \biggl( -\frac{BH(T)}{R~T}\biggr)}_{k_{des}(T)} \; .
\end{equation}
The maximum temperature peak is found by differentiating the above equation as a function of T and setting $d^2N/dT^2 =0$. 
Since both the prefactor $\nu$ and BH are functions of T, the derivative has to be solved numerically by integrating the differential equation with, for example, the Euler method.
In the approximation that $\nu$ and BH are not temperature dependent, the temperature peak is obtained by solving the root of the following equation numerically:
\begin{equation}\label{eq:T_peak_approx}
    \frac{BH}{R ~T^2} - \frac{\nu}{\beta} \exp \biggl( -\frac{BH}{R~T}\biggr) = 0 \; .
\end{equation}

However, Eq.\ref{eq:T_peak_approx} was never used; indeed, BH(T) and v(T) were always calculated according to Eq.\ref{eq:tpd}, i.e. taking into account the temperature dependence.

\section{Results}

In this work we used adsorption of astrochemical representative molecules (water, ammonia, methanol) on both amorphous and crystalline ice models to study the dependency of i) the vibrational partition function on the desorption rate (Subsec.~\ref{subsec:6HRT}), ii) the different prefactor models on simulated TPD spectra (Subsec.~\ref{subsec:nuTPD}) and iii) the thermal correction to the BE (Subsec.~\ref{subsec:bht}). Structural models and binding energies for the amorphous ice were retrieved from our previous papers and are based on large molecular clusters aimed at simulating the high variability of sites of the amorphous ice mantles.\cite{Tinacci_NH3_BE, tinacci_2023_water, bariosco2025methanol} 
New \textit{ad hoc} calculations have been run on the crystalline ice only, simulated within the periodic boundary conditions. Even if not representative of the interstellar ices, it allows to run quick calculations while exhibiting the soft nature imposed by the hydrogen bond web. This, in turn, gives phonon modes which may be perturbed upon adsorption at the surface.
In Table \ref{tab:be} we report the overview of all the computed properties regarding the TPD spectra, \textit{i.e.} binding energies, thermal corrections, and the calculated pre-exponential factors and corresponding T$_{peak}$.

In Appendices \ref{app:be} and \ref{app:comp_details} we report the BE formalism and the methodology used to carry out the simulations, respectively. The accuracy about the chosen methodology for the interested reader is shown in the Supplementary Material.

\begin{table*}
\caption{\label{tab:be} Properties of H$_2$O, NH$_3$ and CH$_3$OH as adsorbed on molecular cluster (Mol) or at the crystalline ice models (Cry).
T$_{peak}$ is extracted from the simulated TPD spectra, using $\beta$ = 0.04 K/s and A = $10^{-19}$ $m^2$. 
The prefactors $\nu_{TST}^{vib}$ and $\nu^{Tait}_{TST}$ are computed at the T$_{peak}$.
$\nu_{HH}$ is computed using BH(0). 
Energy values are in kJ/mol, peak temperatures in Kelvin and prefactors in $s^{-1}$.}
\begin{ruledtabular}
\begin{tabular}{lccccccccccc}
Species & Sample & BH(0) & $\Delta BH\bigl(T_{peak}\bigr)$ & $\nu_{TST}^{vib}\bigl(T_{peak}\bigr)$ & $\nu_{Campbell}\bigl(T_{peak}\bigr)$ & $\nu^{Tait}_{TST}\bigl(T_{peak}\bigr)$ & $\nu_{HH}(BH(0))$ & $T_{peak}^{vib}$ & $T_{peak}^{Campbell}$ & $T_{peak}^{Tait}$ & $T_{peak}^{HH}$ \\
\hline
\multirow{5}{*}{H$_2$O}   & Mol 1                   & 30.9 & 2.7 & 5.9 $\cdot 10^{14}$ & 1.2 $\cdot 10^{14}$  & 1.3 $\cdot 10^{15}$ & 1.9 $\cdot 10^{12}$ & 105 & 110 & 103 & 123 \\
                          & Mol 2                   & 44.6 & 3.0 & 1.4 $\cdot 10^{15}$ & 1.9 $\cdot 10^{14}$  & 4.0 $\cdot 10^{15}$ & 2.2 $\cdot 10^{12}$ & 146 & 153 & 142 & 173 \\
                          & Mol 3                   & 56.6 & 3.8 & 7.0 $\cdot 10^{15}$ & 2.5 $\cdot 10^{14}$  &8.4 $\cdot 10^{15}$ & 2.5 $\cdot 10^{12}$  & 180 & 191 & 179 & 221 \\
                          & Cry Ads & 50.4 & 3.3 & 1.4 $\cdot 10^{15}$ &  2.2 $\cdot 10^{14}$  &5.9 $\cdot 10^{15}$ & 2.4 $\cdot 10^{12}$  & 163 & 171 & 158 & 193 \\
                          & Cry Des & 71.1 & 6.1 & 1.3 $\cdot 10^{16}$ & 3.4 $\cdot 10^{14}$  &1.5 $\cdot 10^{16}$ & 2.8 $\cdot 10^{12}$  & 214 & 241 & 213 & 267 \\
\hline
\multirow{4}{*}{NH$_3$}   & Mol 1                   & 24.3 & 1.3 & 1.2 $\cdot 10^{14}$ & 9.6 $\cdot 10^{13}$  &9.3 $\cdot 10^{14}$ & 1.7 $\cdot 10^{12}$ & 85  & 85 & 81  & 95  \\
                          & Mol 2                   & 34.1 & 2.1 & 2.7 $\cdot 10^{14}$ & 1.5 $\cdot 10^{14}$  &2.8 $\cdot 10^{15}$ & 2.0 $\cdot 10^{12}$ & 116 & 118 & 110 & 132 \\
                          & Mol 3                   & 44.5 & 2.8 & 1.1 $\cdot 10^{15}$ & 2.1 $\cdot 10^{14}$  &6.0 $\cdot 10^{15}$ & 2.3 $\cdot 10^{12}$ & 145 & 151 & 140 & 171 \\
                          & Cry Ads & 55.2 & 2.9 & 1.3 $\cdot 10^{15}$ & 2.8 $\cdot 10^{14}$  &1.2 $\cdot 10^{16}$ & 2.6 $\cdot 10^{12}$ & 177 & 184 & 169 & 208\\
\hline
\multirow{4}{*}{CH$_3$OH} & Mol 1                   & 20.3 & 1.7 & 2.3 $\cdot 10^{16}$ & 2.4 $\cdot 10^{14}$ &2.8 $\cdot 10^{16}$ & 1.1 $\cdot 10^{12}$  & 64  & 72 & 64  & 84  \\
                          & Mol 2                   & 35.9 & 1.8 & 4.7 $\cdot 10^{16}$ & 5.0 $\cdot 10^{15}$  &1.7 $\cdot 10^{17}$ & 1.5 $\cdot 10^{12}$ & 107 & 119 & 104 & 139 \\
                          & Mol 3                   & 55.1 & 1.9 & 8.5 $\cdot 10^{16}$ & 8.7 $\cdot 10^{14}$  &6.7 $\cdot 10^{17}$ & 1.9 $\cdot 10^{12}$ & 158 & 175 & 151 & 206 \\
                          & Cry Ads & 60.1 & 1.4 & 2.8 $\cdot 10^{16}$ & 9.6 $\cdot 10^{14}$  &9.3 $\cdot 10^{17}$ & 2.0 $\cdot 10^{12}$ & 174 & 188 & 162 & 221 \\
\end{tabular}
\end{ruledtabular}
\end{table*}

\subsection{Vibrational partition function and the 6HRT factor}\label{subsec:6HRT}

\begin{figure}
\begin{center}
\subfigure[Water]{
\includegraphics[width=0.99\columnwidth]{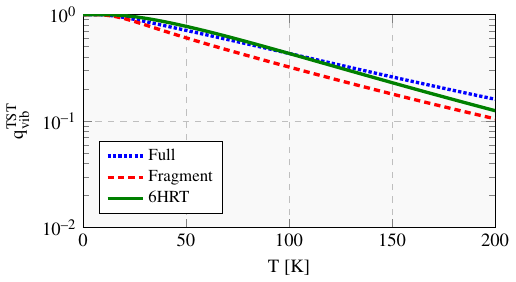}
}
\subfigure[Ammonia]{
\includegraphics[width=0.99\columnwidth]{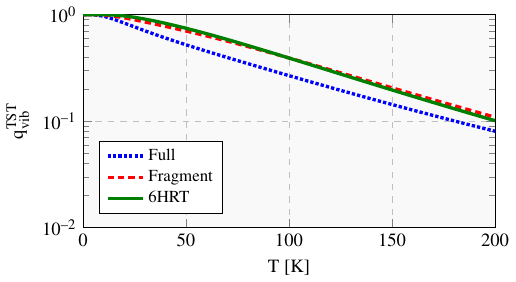}
}
\subfigure[Methanol]{
\includegraphics[width=0.99\columnwidth]{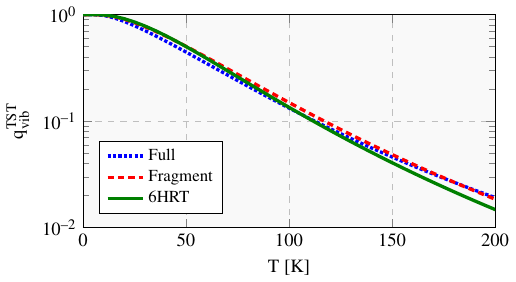}
}
\end{center}
\caption{Vibrational partition function ratio (left member of Eq.~\ref{eq:q_vib_ratio}) of water (a), ammonia (b) and methanol (c) for the crystalline (Cry Ads) cases. 
"Full", "Fragment", and "6HRT" stand for full, and partial (first coordination sphere of the desorbing molecule ("Fragment") and desorbing molecule only ("6HRT")) vibrational frequencies calculations.}
\label{fig:q_ratio}
\end{figure}

The difference between the immobile particle approximation (Eq.~\ref{eq:apprx_pre_exponential}, Tait et al.\cite{tait2005n,tait2006n}) and the full TST approach (Eq. \ref{eq:tst_pre_exponential}) is the $q_{vib}^{TST}$ term decomposed in Eq.~\ref{eq:q_vib_ratio}. 

In the present work, we relied on the vibrational harmonic frequency calculation approximation. 
To verify if the surface phonons play a not negligible contribution to the $q_{vib}^{TST}$ term and the importance of the 6 hindered rotations and translations (6HRT), we calculated the vibrational frequencies on the periodic systems at three levels: i) the whole system ("Full"), ii) the first coordination shell of the desorbed species ("Fragment", described in Appendix~\ref{app:comp_details}), iii) the adsorbed molecule only, using for the bare surface only its electronic energy ("6HRT Mol. Fragment"). 
In Fig.~\ref{fig:q_ratio}, the $q_{vib}^{TST}$ quantity is only presented for the crystalline (Cry) samples. This analysis was not performed on the molecular samples, due to the 2-layers ONIOM approach and to the constraints applied to the geometry optimization (fixing of large part of the ice) which could bias this fine analysis. 
Looking at all the crystalline samples of Fig.~\ref{fig:q_ratio}, the thermal correction to the $q_{vib}^{TST}$ is negligible at low temperatures, as expected. At values close to the desorption temperature the deviation between $\nu^{Tait}_{TST}$ and $\nu_{TST}^{vib}$ is 2 orders of magnitude (methanol), which however corresponds to a modest difference in the T$_{peak}$ (12 K).
The term $q_{vib}^{TST}(M)$ was also studied, even if not presented in Fig.~\ref{fig:q_ratio}, and for all the cases it is always $\approx 1$, meaning that the intramolecular vibrations of the adsorbed and isolated molecule are never excited even at high temperatures, as expected due to the absence of very small vibrational frequencies for the adopted molecules (the lowest one being $\sim$300 cm$^{-1}$ for methanol in the gas phase).

As regards the calculation of vibrational frequencies performed at the three levels, the $1/q_{vib}^{6HRT}$ (6HRT Mol. Fragment) gives the same results of the $q_{vib}^{TST}$ calculated with a complete ("Full") and partial ("Fragment") vibrational analysis. This means that surface phonons do not have an important contribution to $q_{vib}^{TST}$, but it is enough to consider the partition function of the detaching molecule only.
Therefore, we can conclude that, even for systems where the coupling between vibrational frequencies of the adsorbate and the adsorbent is relevant, the calculations of $q_{vib}^{6HRT}$ is already enough to ensure good accuracy.
This is particularly important to save computational resources, because of the high cost of performing a vibrational calculation on the full system.

\subsection{Pre-exponential factor impact on TPD spectra}
\label{subsec:nuTPD}

\begin{figure}
\begin{center}
\subfigure[Water Mol]{
\includegraphics[width=0.99\columnwidth]{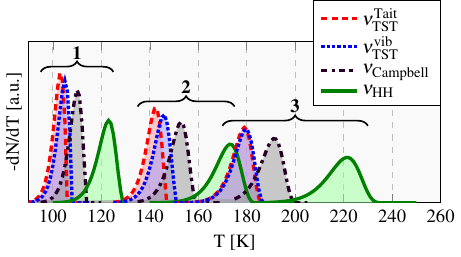}
}
\subfigure[Ammonia Mol]{
\includegraphics[width=0.99\columnwidth]{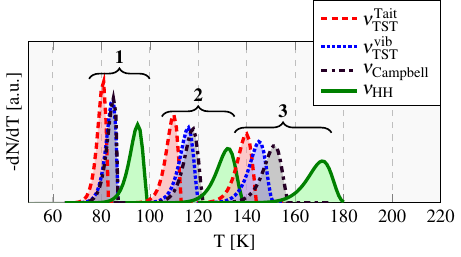}
}
\subfigure[Methanol Mol]{
\includegraphics[width=0.99\columnwidth]{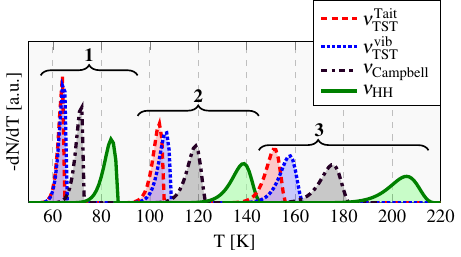}
}
\end{center}
\caption{Simulated TPD spectra of molecular (Mol) samples of water (a), ammonia (b) and methanol (c) (see Appendix \ref{app:comp_mol}). 
The spectra are computed as first-order desorption and heating rate $\beta = 0.04 K/s$ and A = $10^{-19}$ $m^2$. 
Each sample is identified by the number above the curly bracket.}
\label{fig:TPD_cluster}
\end{figure}

\begin{figure*}
\begin{center}
\subfigure[Water Cry Ads]{
\includegraphics[width=0.99\columnwidth]{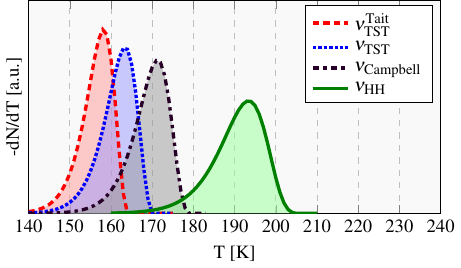}
}
\subfigure[Water Cry Des\label{subfig:cry_des}]{
\includegraphics[width=0.99\columnwidth]{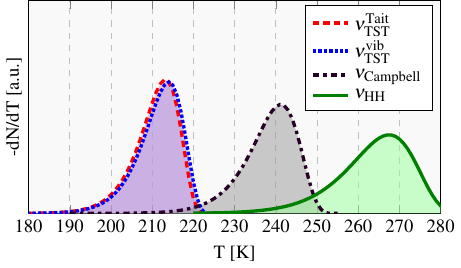}
}
\subfigure[Ammonia Cry Ads]{
\includegraphics[width=0.99\columnwidth]{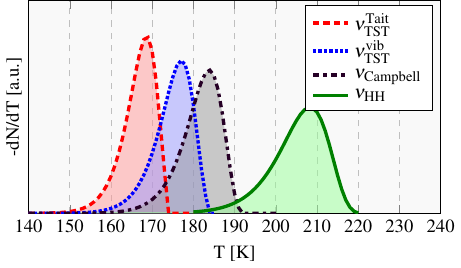}
}
\subfigure[Methanol Cry Ads]{
\includegraphics[width=0.99\columnwidth]{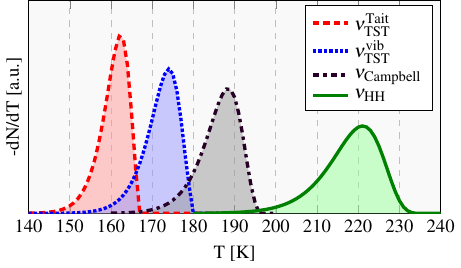}
}
\end{center}
\caption{Simulated TPD spectra of the crystalline (Cry) samples (water (a), ammonia (b) and methanol (c) see Appendix \ref{app:comp_periodic}). 
The spectra are computed as first-order desorption and heating rate $\beta$ = 0.04 $K/s$ and A = $10^{-19}$ $m^2$.}
\label{fig:TPD_cry}
\end{figure*}

For each sample, the desorption rate was calculated with the four different prefactor models discussed before: $\nu_{TST}^{vib}$ (Eq.~\ref{eq:tst_pre_exponential}), $\nu^{Tait}_{TST}$ (Eq.~\ref{eq:apprx_pre_exponential}, proposed by Tait et al. \cite{tait2005n,tait2006n}), $\nu_{HH}$ (Eq.~\ref{eq:HH} proposed by Hasegawa \& Herbst \citep{Hasegawa1992}) and $\nu_{Campbell}$ (Eq.~\ref{eq:prefactor_campebel} proposed by Campbell \& Sellers \cite{campbell2012entropies}). 
The temperature peaks of simulated TPDs were calculated according to Eq.~\ref{eq:tpd}, \textit{i.e.} assuming no temperature dependency on BH(T) and $\nu (T)$, and the simulated TPD spectra are reported in Figs. \ref{fig:TPD_cluster} and \ref{fig:TPD_cry} for the molecular and periodic samples, respectively.
Table \ref{tab:be} summarises the major parameters of Eq.~\ref{eq:tpd}.

As a general trend, the order of the different prefactors is
\begin{equation*}
    \nu_{HH} < \nu_{Campbell} < \nu_{TST}^{vib} < \nu^{Tait}_{TST}
\end{equation*}
and, accordingly, the opposite for the $T_{peak}$, \textit{i.e.}
\begin{equation*}
    T_{peak}^{Tait} < T_{peak}^{vib} < T_{peak}^{Campbell} < T_{peak}^{HH} \; .
\end{equation*}
For all the computed species the smallest prefactor is $\nu_{HH}$, which is 3 to 5 orders of magnitude lower than $\nu^{Tait}_{TST}$, which, as pointed out in Section \ref{sec:rate}, represents an upper limit to the prefactor, as it relies on the immobile particle approximation.\cite{tait2005n,tait2006n} $\nu_{Campbell}$ is 1-2 orders of magnitude lower than $\nu^{Tait}_{TST}$ (which agrees with the previous findings by Tait et al.\cite{tait2005n,tait2006n}) and the full TST approach ($\nu_{TST}^{vib}$) stays in the middle; depending on the sample it can randomly collapse to the $\nu^{Tait}_{TST}$ or the $\nu_{Campbell}$ solutions.
Considering the various approximations affecting the prediction of the BE with the adopted icy models and the limits in accuracy of the DFT theory, we believe that adopting either $\nu^{Tait}_{TST}$ or $\nu_{Campbell}$ represent the best cost-accuracy trade-off (with a small difference on the $T_{peak}$ of at most 30 K) considering how easy is their evaluation, free from expensive quantum mechanical calculations needed for the $\nu_{TST}^{vib}$. In the Supplementary Material we report the analysis of the different prefactors on CH$_4$ and CH$_3$OH adsorbed on MgO (100) surface, showing that $\nu_{TST}^{vib}$ correctly reproduces the experimental results by Campbell et al.\cite{campbell2012entropies}

The large discrepancy of $\nu_{HH}$ can be attributed to the fact that, in contrast to the TST approach, it does not take into account the rotational motions, whose moment of inertia increases with the molecular size.\cite{tait2005n,tait2006n} Indeed the deviation of $\nu_{HH}$ and $T_{peak}^{HH}$ with respect to the others pre-factors notably increases for methanol.

Another general trend is that the stronger the binding energy, the larger the discrepancy among the different methods for calculating the $T_{peak}$, as can be graphically seen by the length of the curly parenthesis of Fig. \ref{fig:TPD_cluster}.

Finally, we want to highlight the effect of the $\beta$ factor in the simulated TPD experiments. 
Lower $\beta$s correspond to narrower peaks and lower $T_{peak}$, and \textit{viceversa}. 
Therefore, when $\beta$ decreases the difference in simulated TPDs using different prefactor models will be minimized. 
This is expected at least for the TPD computed with $\nu^{Tait}_{TST}$ and $\nu_{TST}^{vib}$.

\subsection{Desorption rate Temperature dependencies}
\label{subsec:bht}

In computational astrochemistry, it is common to report the BE corrected only for the ZPE without any thermal correction (see Appendix \ref{app:be}). This approximation makes sense when temperatures are very low, such as in cold molecular clouds ($\approx$ 10 K), but at higher temperatures it may become not negligible. 
Indeed, using the samples described in Appendix \ref{app:comp_details}, we computed the thermal corrections to the BE at the T$_{peak}$. For the majority of the cases they are in the range of 1.5--4 kJ/mol, representing $\approx 5 \%$ of the total BE, which is a lower contribution with respect to the ZPE ($\approx 20 \%$ of the BE), but somehow important.
We also studied the differences between the full rigid rotor harmonic oscillator (RRHO) and quasi-RRHO (q-RRHO) correction to the BH(0). The discrepancy is always positive, meaning that $\Delta$BH(T)(RRHO) is always larger than $\Delta$BH(T)(q-RRHO), and very small $\approx$ 0--1 kJ/mol, below the typical desorption temperature.
Due to the low correction of q-RRHO with respect to RRHO at our operating temperatures and the fact that q-RRHO theory was developed to correct enthalpies at higher temperatures, we suggest, if needed, to use the normal RRHO to correct the BH(0). See Appendix \ref{app:q_vib_appendix} for details.

In the majority of the simulated cases, the assumption of modelling $k_{des}(T)$ with Eq.~\ref{eq:k_des}, with $\nu$ and the BH calculated at the $T_{peak}$, \textit{i.e}:
\begin{equation}
    k_{des}(T) = \nu_{TST}^{vib}(T_{peak}) \exp \Biggl(-\frac{BH(T_{peak})}{R~ T}\Biggr) \; ,
\end{equation}
describes with very good accuracy the desorption process. 
In fact, as shown in the $\Delta BH(T_{peak})$ column of Table~\ref{tab:be}, the thermal corrections to the BH(0) remain constant in the 50--300 K range for almost all the samples. 
On the other hand, the prefactor mostly changes (one order of magnitude) in the 50--200 K range for all the considered samples (see $\nu_{TST}^{vib}(T_{peak})$ column in Table~\ref{tab:be} for different samples). 
This behavior is mainly due to the importance of T in the exponential factor of $k_{des}(T)$, overcoming the thermal correction to the BH(0) and to the prefactor. 
In an experiment involving a multilayer regime of ethane, Luna et al.\cite{luna2015experimental} experimentally found that the prefactor depends on the temperature in a desorption mechanism described by the Polanyi-Wigner equation, thus supporting the above discussion.

To correctly estimate the dependence of $k_{des}(T)$ (Eq. \ref{eq:k_des}) on the temperature we relied on the three-parameter modified Arrhenius equation
proposed by Kooij \citep{kooij1893zersetzung,laidler1996glossary} (as commonly used by the astro-chemical community for fitting reaction rate in gas phase): 
\begin{equation}\label{eq:kooij}
k_{des}^{Kooij}(T) = \alpha \biggl(\frac{T}{300}\biggr)^\eta \exp \biggl( -\frac{\gamma}{R~T}\biggr) \; ,
\end{equation}
where $\alpha$, $\eta$, and $\gamma$ are the fitting parameters, which represent the pre-exponential factor, a corrective temperature dependent term to the successful collisions of the pre-exponential factor, and the activation energy (in our case BH(T)), respectively.
As one can see, for $\eta = 0$, the corrective term $(\frac{T}{300})^\eta$ is equal to 1 and, accordingly, we obtain the Arrhenius equation. By applying the same procedure as for Eq.\ref{eq:tpd}, the temperature peak (T$_{peak}$) is found imposing the second derivative of T equal to zero ($d^2N/dT^2 =0$), thus obtaining:

\begin{equation}\label{eq:t_peak_kooij}
    \frac{\gamma}{R} + \eta~ T -  \frac{T^2}{\beta} \alpha \biggl(\frac{T}{300}\biggr)^\eta \exp \biggl( -\frac{\gamma}{R~T}\biggr) = 0 \; .
\end{equation}
This equation can be easily solved with a numerical algorithm, \textit{e.g.} bisection or Newton. 
The fitting procedure can be applied to whatever TPD spectrum to derive a posteriori the activation energy ($\gamma$, \textit{i.e.}$E_a$) and the pre-exponential factor ($\alpha$, \textit{i.e.} $\nu$).

In Appendix~\ref{app:kooij}, we demonstrate the accuracy obtained when the Arrhenius-Kooij equation is used to fit the desorption rate and also the accuracy of the Eq.~\ref{eq:t_peak_kooij} in the $T_{peak}$ calculation. Briefly, the computed TPD confirm that the fitted rate closely matches the real one, with \( T_{peak} \) differences below 1 K.

\section{Conclusions}
 
In this work, we study the importance of the pre-exponential factor in the desorption rate of some astrochemical representative species (water, ammonia, methanol) from interstellar icy mantles, by means of periodic and molecular DFT simulations. 

Specifically, we show the effect of using different approaches to calculate the prefactor, from the most used one in the astrochemical community ($\nu_{HH}$) to others that explicitly take into account the partition functions or entropy of the involved species ($\nu_{TST}^{vib}$, $\nu^{Tait}_{TST}$ and $\nu_{Campbell}$). 
As expected, for all the samples, either molecular or periodic, $\nu_{HH}$ leads to the typical values discussed in the introduction ($\sim 10^{12}$ s$^{-1}$), while $\nu_{TST}^{vib}$ and $\nu^{Tait}_{TST}$ may differ from 2 up to 5/6 orders of magnitude. 
These differences largely affect the simulated TPD spectra; typically, $\nu_{HH}$ and $\nu^{Tait}_{TST}$ always represent the upper and lower limits on the range of desorption temperature, respectively. 
The difference in the $\mathrm{T_{peak}}$ adopting one approach or another spans from 20 to 60 K, depending on the cases and experimental conditions. 
According to our results, we can say that a more accurate upper limit to the desorption temperature (\textit{i.e.} a lower limit to the prefactor) is $\nu_{Campbell}$, which reduce of a factor of 2-3 the error given by $\nu_{HH}$.
Therefore, our recommendation is to use $\nu^{Tait}_{TST}$ and $\nu_{Campbell}$ to obtain a minimal error to prefactor and desorption temperature; this means that the theories developed by Tait and Campbell can have a more general applicability than what demonstrated so far, \textit{i.e.} they work reasonably well even for H-bonded systems (\textit{i.e.} for "soft" surfaces, like molecular crystals), and not only for van der Waals systems (\textit{i.e.} for "hard" surfaces, i.e. metallic, metal oxides, zeolites, etc.). In case of prefactors estimated throughout quantum mechanical simulations, we have demonstrated that $\nu_{TST}^{vib}$ can be calculated with a very good accuracy only considering the partition functions of the detaching molecule ($1/q_{vib}^{6HRT}$), even for systems in which the coupling of the vibrational modes between adsorbate/adsorbent is large as in the present case. This means that it is enough to calculate a small portion of the total hessian matrix of the system, thus saving large part of the computational effort.

We also show how to easily implement temperature effects in kinetic models through the Kooij-Arrehnius fitting of the desorption rate, $k_{des}(T)$, obtaining accurate values of desorption temperature, activation energy and pre-exponential factor.

\section*{Online scripts and database}

The python scripts, with all the examples, can be found at the following GitHub link: \textcolor{blue}{\url{https://github.com/TinacciL/BE_prefactor}}. 
To easily handle the data set of BE samples (atomic coordinates and energy values summarized in Table \ref{tab:be}), we developed and made publicly available a website based on the molecule hyperactive JSmol plugin (Jmol: an open-source Java viewer for chemical structures in 3D\footnote{\textcolor{blue}{\url{http://www.jmol.org/}}}). This extended electronic version of the calculated structures is available at: \textcolor{blue}{\url{https://tinaccil.github.io/Jmol_BE_prefactor_visualization/}}.

\section*{Supplementary Material}
In the Supplementary Material available online we provide additional information about the accuracy of the DFT method chosen, and on the surface modeling and relative convergence of the binding energy. Moreover, prefactors and TPDs for CH$_4$ and CH$_3$OH adsorbed on MgO (100) are calculated with all the models, in order to check the accuracy of the $\nu_{TST}^{vib}$ in comparison with $\nu_{Campbell}$.

\begin{acknowledgments}
This project has received funding within the European Union’s Horizon 2020 research and innovation programme from the European Research Council (ERC) for the projects “The Dawn of Organic Chemistry” (DOC), grant agreement No 741002, “Quantum Chemistry on Interstellar Grains“ (QUANTUMGRAIN), grant agreement No. 865657, and from the Marie Sk{\l}odowska-Curie for the project ”Astro-Chemical Origins” (ACO), grant agreement No 811312.
PU and SP acknowledge the Italian Space Agency for co-funding the Life in Space Project (ASI N. 2019-3-U.O).
This research has received funding from the Project CH4.0 under the MUR program “Dipartimenti di Eccellenza 2023-2027” (CUP: D13C22003520001).
A. R. acknowledges Accademia delle Scienze di Torino for supporting the project "In silico interstellar grain-surface chemistry". A.R. gratefully acknowledges support through 2023 ICREA Award.
We acknowledge the EuroHPC Joint Undertaking for awarding this project access to the EuroHPC supercomputer LUMI, hosted by CSC (Finland) and the LUMI consortium through a EuroHPC Regular Access call.
F. Dulieu and P. Theulé are thanks of the insight on the laboratory point of view.
LT is grateful to Jacopo Lupi, Leonardo Miele, Franciele Kruczkiewicz and Stefano Ferrero for insightful discussions.
\end{acknowledgments}

\section*{References}

\bibliography{aipsamp}

\newpage

\clearpage

\appendix

\section{Formalism to derive the binding energy}\label{app:be}

The equation adopted for the calculation of the BEs is:
\begin{equation}\label{eq:BE}
\mathrm{BE} = - \Delta \mathrm{E} = \mathrm{E}^{iso}_{ads} + \mathrm{E}^{iso}_{srf} - \mathrm{E}_{c} \; ,
\end{equation}
where E$_c$ is the energy of the complex and E$^{iso}$ the energies of the isolated systems, with "ads" standing for the adsorbate, and "srf" for the surface. 
The BE of the complex is also corrected for the Basis Set Superposition Error (BSSE),\cite{Counterpoise} if not already taken into account by definition in the chosen method.

BE can be decomposed in the pure electronic interaction ($\mathrm{BE}_e$) and the deformation energy ($\mathrm{\delta E}_{def}$) contributions. The $\mathrm{BE}_e$ is given by:
\begin{equation}\label{eq:BSSE}
    \mathrm{BE}_{e} = \mathrm{E}^{iso//c}_{ads}+ \mathrm{E}^{iso//c}_{srf} - \mathrm{E}_{c} \; ,
\end{equation}
where $\mathrm{E}^{iso//c}_{ads}$ and $\mathrm{E}^{iso//c}_{srf}$ are the energies of the isolated adsorbate and the grain at the geometry of the complex ($iso//c$). The total deformation energy ($\mathrm{\delta E}_{def}$) is defined as:
\begin{equation}
    \mathrm{\delta E}_{def} = \underbrace{\bigl( \mathrm{E}^{iso//c}_{ads}-\mathrm{E}^{iso}_{ads}\bigr)}_{\mathrm{\delta E}^{ads}_{def}} +\underbrace{\bigl( \mathrm{E}^{iso//c}_{srf}-\mathrm{E}^{iso}_{srf}\bigl)}_{\mathrm{\delta E}^{srf}_{def}} \; ,
\end{equation}
where $\mathrm{\delta E}^{ads}_{def}$ and $\mathrm{\delta E}^{srf}_{def}$ are the deformation energies of the adsorbate and the surface, respectively.
$\mathrm{\delta E}_{def}$ is therefore expected to be a positive quantity, even if exceptions sometimes occur (for a detailed discussion, please refer to Ref. \onlinecite{Tinacci_NH3_BE}).

In addition, vibrational frequencies are computed to obtain the zero-point energies (ZPE), from which the $\Delta$ZPE results as:
\begin{equation}
    \Delta \mathrm{ZPE} = \mathrm{ZPE}^{iso}_{ads} + \mathrm{ZPE}^{iso}_{srf} - \mathrm{ZPE}_{c} \; .
\end{equation}

Including all the above-mentioned contributions, Eq.~\ref{eq:BE} becomes:
\begin{equation}
    \label{eq:BE_decompose}
    \mathrm{BH(0)} = \underbrace{\mathrm{BE}_e - \mathrm{\delta E}_{def} }_{\mathrm{BE}} + \Delta \mathrm{ZPE} \; .
\end{equation}
If the thermal correction is taken into account the BE at a given temperature T is:
\begin{equation}
    \label{eq:BE_thermal}
    \mathrm{BH(T)} = \mathrm{BH(0)} + \mathrm{4RT} + \mathrm{H_{ads}^{vib}(T)} + \mathrm{H_{srf}^{vib}(T)} - \mathrm{H_{c}^{vib}(T)}  \; ,
\end{equation}
where the 4RT term (R is the ideal gas constant and T the temperature) comes from the classical rotational (3/2RT) and translational (3/2RT) contributions of the isolated adsorbate molecule (non-linear in the selected cases) in the rigid rotor approximation, and 1RT comes from the volume work contribution to the enthalpy. 
The H$^{vib}$(T) terms refer to the thermal contribution to the enthalpy without the ZPE already taken into account in Eq.~\ref{eq:BE_decompose}, that in the case of rigid rotor harmonic oscillator (RRHO) is:
\begin{equation}\label{eq:rrho}
    \mathrm{H}^{\mathrm{vib}}_{RRHO}(\mathrm{T}) = R \sum_i^{3N-6} \frac{h ~\nu_i / k_B}{\exp \Bigl( \frac{h ~\nu_i}{k_B ~T} \Bigr)-1} \; ,
\end{equation}
where h is the Planck constant, $k_B$ the Boltzmann constant, $\nu$ the normal mode vibrational frequency and N the number of atoms of the system.

The vibrational enthalpy ($\mathrm{H}^{\mathrm{vib}}(\mathrm{T})$) can also be corrected as proposed by Martin Head-Gordon's group,\cite{li2015improved} where the rigid rotor-harmonic oscillator approximation can be improved by treating low-lying vibrational modes as free translational and rotational modes via a quasi-RRHO (q-RRHO) model:
\begin{equation}\label{eq:quasi_rrho}
    \mathrm{H}^{\mathrm{vib}}_{q-RRHO}(\mathrm{T}) = R \sum_i^{3N-6} \omega(\nu_i) \frac{h ~\nu_i / k_B}{\exp \Bigl( \frac{h ~\nu_i}{k_B ~T} \Bigr)-1} + T \frac{1 - \omega(\nu_i)}{2}\; ,
\end{equation}
where $\omega(\nu_i)$ is defined as follows:
\begin{equation}\label{eq:omega}
    \omega(\nu_i) = \frac{1}{1 + (\nu_0/\nu_i)^4} \; ,
\end{equation}
and $\nu_0$ is the threshold frequency for considering the vibrational modes as free translations and rotations ($<\nu_0$) or harmonic vibrations ( $>\nu_0$).
In the present work, $\nu_0$ has been set to 100 $cm^{-1}$.

\section{Campbell prefactor}\label{app:campbell}

Campbell et al. \cite{campbell2012entropies,campbell2013kinetic} demonstrated that, in the case of methanol adsorbed on Pt surface, the adoption of the harmonic oscillator approximation could lead to significant deviations of the prefactor from experimental values. \cite{campbell2013kinetic} 
Moreover, they found that the standard entropies of the adsorbed molecules linearly depend on the entropy of the gas-phase molecule at the same T, \textit{i.e.} S$^\circ_{gas}$,\cite{campbell2012entropies} with the equation:
\begin{equation}
    S^\circ_{ad}= 0.70 S^\circ_{gas} -3.3 R.
    \label{eqn:linear_campbell_gas_entro}
\end{equation}
The entropy of the transition state for desorption (S$^\circ_{TS,des}$) is identical to S$^\circ_{gas}$ at the same temperature,\cite{campbell2013kinetic,madix1979preexponential} except for the missing translational degree of freedom, thus:
\begin{equation}
    S^\circ_{TS,des}= S^\circ_{gas} - S^\circ_{gas,1D-trans} ,
    \label{eqn:ts_des_entropy}
\end{equation}
where S$^0_{gas,1D-trans}$ for any gas, can be easily computed using the Sackur-Tetrode equation:\cite{mcquarrie2008quantum}
\begin{equation}
    S^\circ_{gas,1D-trans} = \frac{1}{3} \left\{ S^\circ_{Ar,298K} + R \ln{\left[\left(\frac{m_A}{m_{Ar}}\right)^{\frac{3}{2}} \left(\frac{T}{298K}\right)^{\frac{5}{2}}\right]} \right\},
\end{equation}
where m$_A$ is the molar mass of the gas, m$_{Ar}$ is the molar mass of the Argon gas, and S$^\circ_{Ar,298K}$ is the entropy of Ar gas at 1 bar and 298K (=18.6R). 
Expressing the Polanyi-Wigner equation (Eq.\ref{eq:WP}) in terms of S$^\circ_{TS,des}$ and S$^0_{ad}$ gives:\cite{atkins2023atkins}
\begin{equation}
    \nu_{des} = \frac{k_B T}{h} \exp\left({\frac{S^0_{TS,des} - S^0_{ad}}{R}}\right).
    \label{eqn:prefactor_funct_ts_entro}
\end{equation}
Now, by substituting Eq. \ref{eqn:linear_campbell_gas_entro} and Eq. \ref{eqn:ts_des_entropy} into Eq. \ref{eqn:prefactor_funct_ts_entro}, we obtain the $\nu_{Campbell}$:
\begin{equation}
\begin{split}
    \nu_{Campbell} = &\frac{k_B T}{h} \exp \Bigg\{ 
    \frac{0.30 S^\circ_{gas}}{R} + 3.3 -\\
    &-\frac{1}{3} \Bigg[ 18.6 + \ln \bigg( \bigg(\frac{m_A}{m_{Ar}}\bigg)^{\frac{3}{2}} 
    \bigg(\frac{T}{298\,\text{K}}\bigg)^{\frac{5}{2}} \bigg) \Bigg] 
    \Bigg\}
\end{split}
\end{equation}
The S$^0_{gas}$ can be readily obtained from standard thermodynamic tables; in case of non-listed temperatures it is also possible to extrapolate them by means of the heat capacity.\cite{linstorm1998nist}

\begin{figure}
\includegraphics[width=0.99\columnwidth]{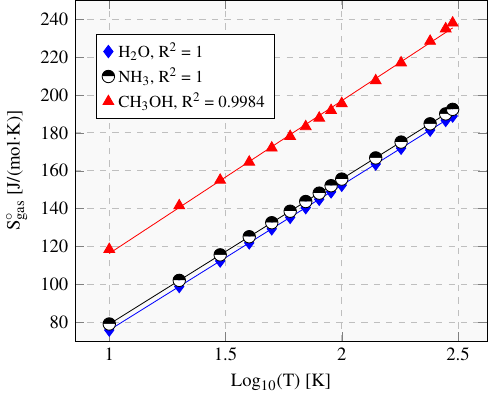}
\caption{Correlation plot between S$^0_{gas}$ and the logarithm of the temperature for H$_2$O, NH$_3$ and CH$_3$OH (Eqs. \ref{eq:campbel_h20}, \ref{eq:campbel_nh3} and, \ref{eq:campbel_ch30h} respectively).}
\label{fig:s0_gas_campbell}
\end{figure}

In this work we propose a new simple way to interpolate the S$^0_{gas}$ without taking advantage of the heat capacity but keeping the same accuracy. 
We calculated the \( S^0_{gas} \) values for the species in the gas phase over the temperature range of 10 K to 300 K using the B97-3c level of theory \cite{b97_3c}. 
The temperature dependence was then linearized to determine the slope, enabling interpolation of \( S^0_{gas} \) values at other temperatures, as shown in Fig. \ref{fig:s0_gas_campbell}. 
This is also important when tabulated values are not available at all temperatures, in particular at the extremely low temperature conditions of cold astronomical environments.
Fig. \ref{fig:s0_gas_campbell} shows the plot of the S$^0_{gas}$ computed at different T for CH$_3$OH, H$_2$O and NH$_3$. 
As illustrated, the data are perfectly described by a linear fit for all the considered molecules, with a small deviation observed just for the methanol case ($R^2$ = 0.9984).
The fitting equations used to interpolate the S$^0_{gas}$ for the different species are the following:
\begin{align}
     S^0_{gas} (H_2O)    &= 76.6 \log_{10}(T) - 0.82 \label{eq:campbel_h20} \\
     S^0_{gas} (NH_3)    &= 76.7 \log_{10}(T) + 2.2\label{eq:campbel_nh3} \\
     S^0_{gas} (CH_3OH)  &= 80.7 \log_{10}(T)  + 35.6 \label{eq:campbel_ch30h}
\end{align}
The prediction accuracy can be tested by interpolating the value of S$^0_{gas}$ at 298K and comparing it with the one reported experimentally in literature.\cite{linstorm1998nist} As shown in Table \ref{tab:fitt_vs_exp_entropy_campbell}, S$^0_{gas}$ for H$_2$O and NH$_3$ are in excellent agreement with the tabulated experimental data. On the contrary, a slight discrepancy of $\sim 5$ J/(mol$\cdot$K) is observed for CH$_3$OH. This small deviation can be attributed to the lower quality of the fit for CH$_3$OH ($R^2$ = 0.9984) with respect to the other molecules ($R^2$ = 1).

\begin{table}
\caption{ S$^0_{gas}$ values (in J/(mol$\cdot$K)) obtained at 298 K from experimental works \cite{linstorm1998nist} and using the fit obtained in this work (Eqs. \ref{eq:campbel_h20}, \ref{eq:campbel_nh3} and \ref{eq:campbel_ch30h}).}
\centering
    \begin{tabular}{c c c c }
    \hline
        \hline
         Molecule &  Experimental & Fitted & Eq. \\
         \hline
         H$_2$O & 188.84  & 188.67 & \ref{eq:campbel_h20} \\
         NH$_3$ & 192.77 & 192.02 & \ref{eq:campbel_nh3} \\
         CH$_3$OH  & 240 & 235.24 & \ref{eq:campbel_ch30h}\\
         \hline
    \hline
    \end{tabular}
    \label{tab:fitt_vs_exp_entropy_campbell}
\end{table}

\section{Computational Methods \& Models}\label{app:comp_details}

In the present work two paradigms were used to simulate realistic and macroscopic water icy grain models: the cluster approach in which the ONIOM (Our own N-layered Integrated molecular Orbital and Molecular mechanics) procedure is exploited to deal with large structures, and the periodic approach to simulate a macroscopic crystalline surface.

\subsection{Molecular calculations}\label{app:comp_mol}

\begin{figure*}
\begin{center}
\subfigure[H$_2$O Mol 1.]{
\includegraphics[height=0.40\columnwidth]{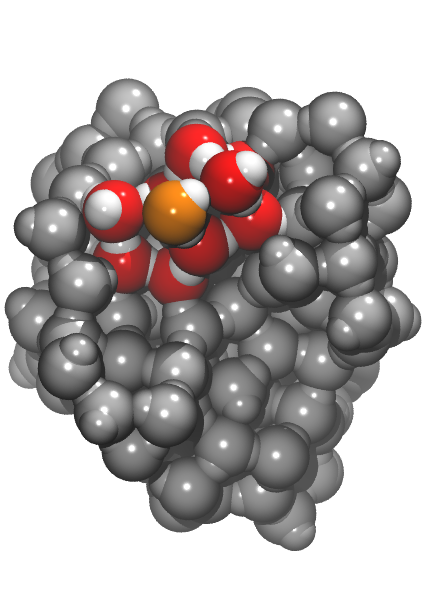}
}
\subfigure[H$_2$O Mol 2.]{
\includegraphics[height=0.40\columnwidth]{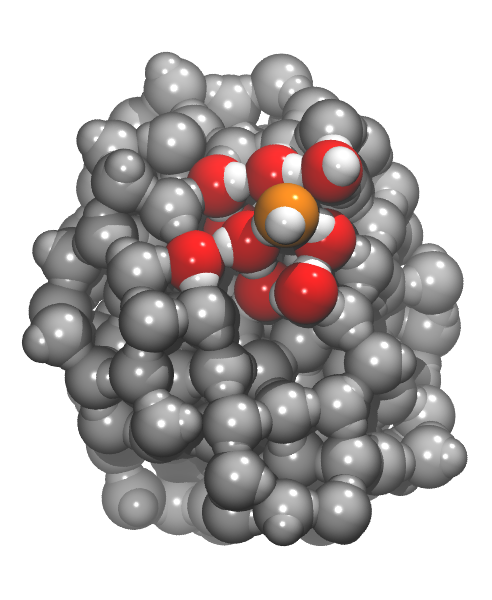}
}
\subfigure[H$_2$O Mol 3.]{
\includegraphics[height=0.40\columnwidth]{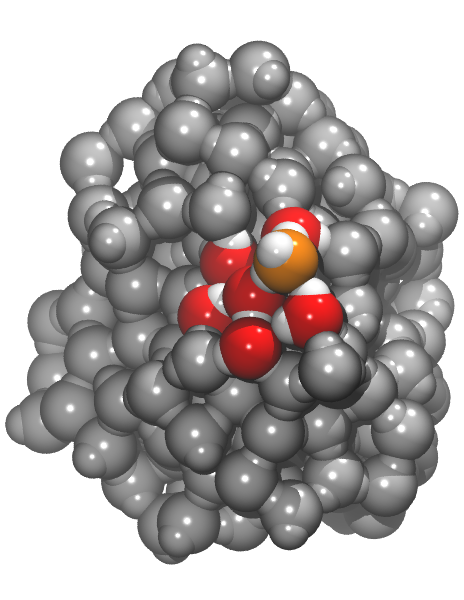}
}
\hfill
\subfigure[NH$_3$ Mol 1.]{
\includegraphics[height=0.40\columnwidth]{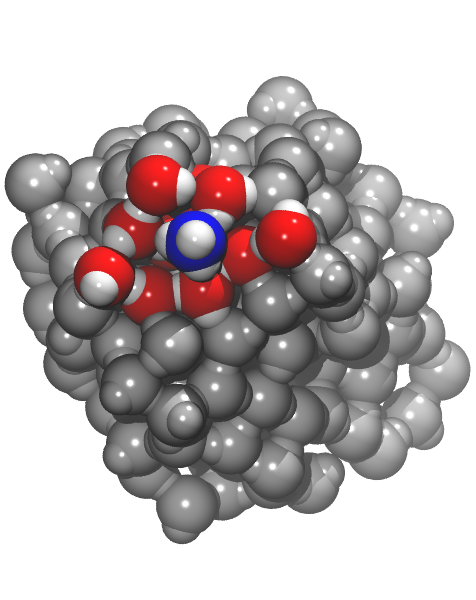}
}
\subfigure[NH$_3$ Mol 2.]{
\includegraphics[height=0.40\columnwidth]{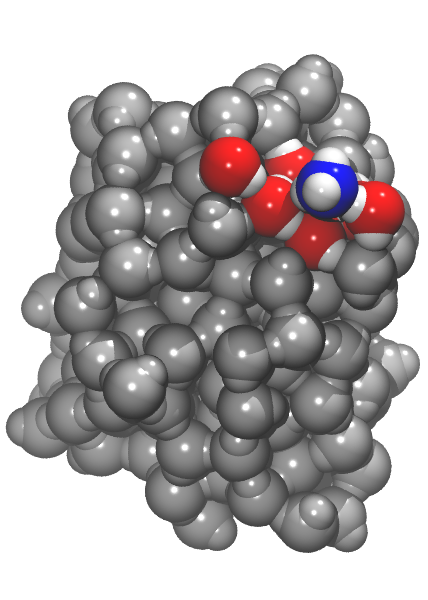}
}
\subfigure[NH$_3$ Mol 3.]{
\includegraphics[height=0.40\columnwidth]{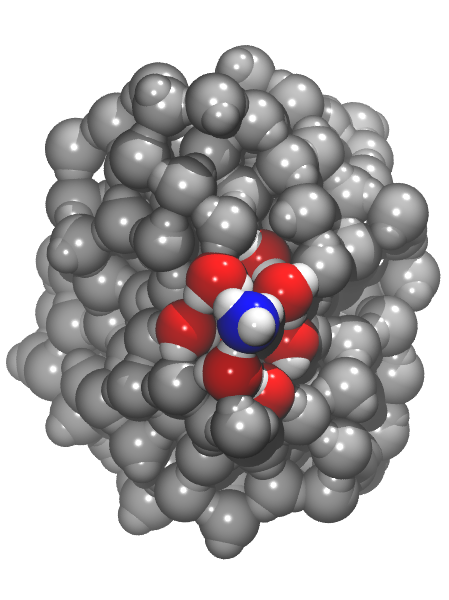}
}\\
\subfigure[CH$_3$OH Mol 1.]{
\includegraphics[height=0.40\columnwidth]{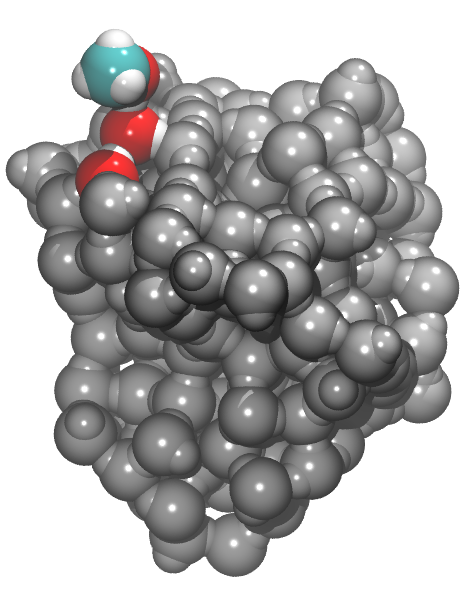}
}
\subfigure[CH$_3$OH Mol 2.]{
\includegraphics[height=0.40\columnwidth]{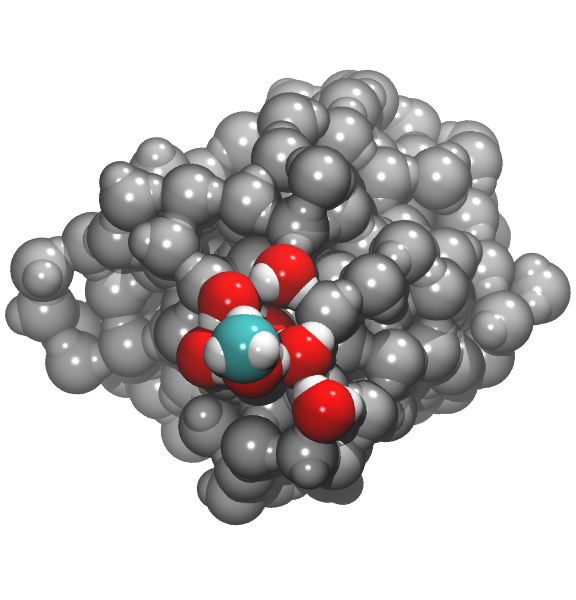}
}
\subfigure[CH$_3$OH Mol 3.]{
\includegraphics[height=0.40\columnwidth]{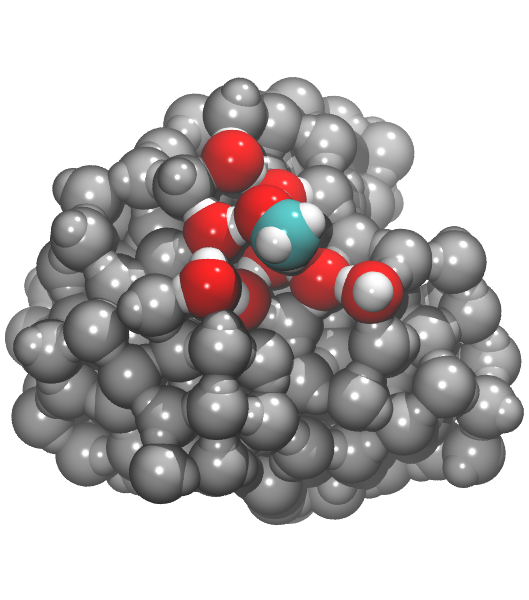}
}
\end{center}
\caption{Optimized geometries of the molecular (Mol) samples: water (a-c), ammonia (d-f) and methanol (g-i). 
The atoms in the ONIOM high-level zone are shown in colors while the ONIOM low-level zone is in gray. 
Atom color legend: oxygen in red, hydrogen in white, nitrogen in blue and carbon in cyan. 
In subfigures (a), (b) and (c) the oxygen of the desorbing water molecule is highlighted in orange.}
\label{fig:mol_samples}
\end{figure*}

\begin{figure*}
\begin{center}
\subfigure[Water Cry Ads.]{
\includegraphics[width=0.15\columnwidth]{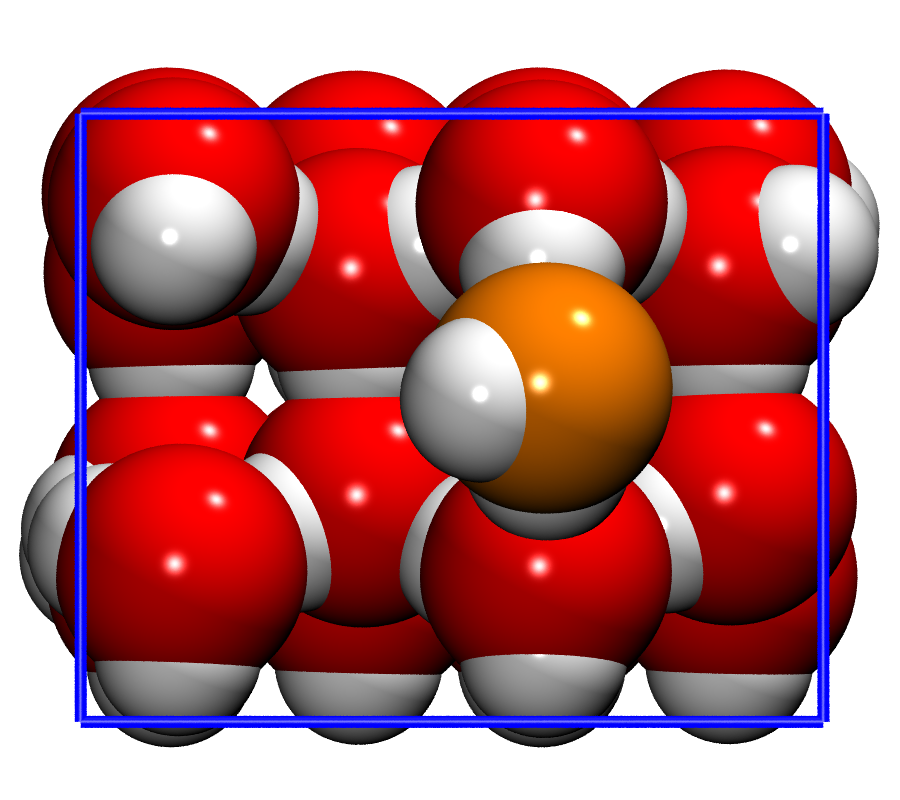}
\includegraphics[width=0.145\columnwidth]{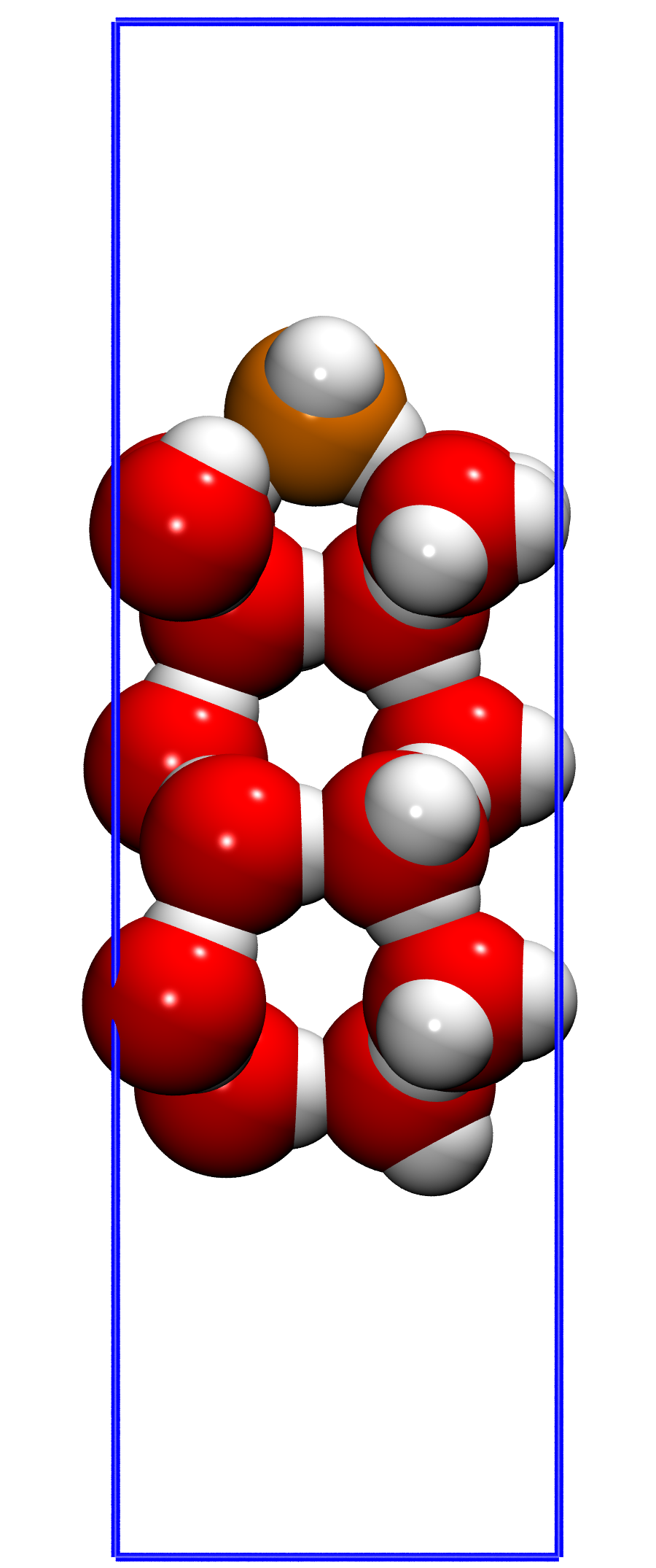}
\includegraphics[width=0.145\columnwidth]{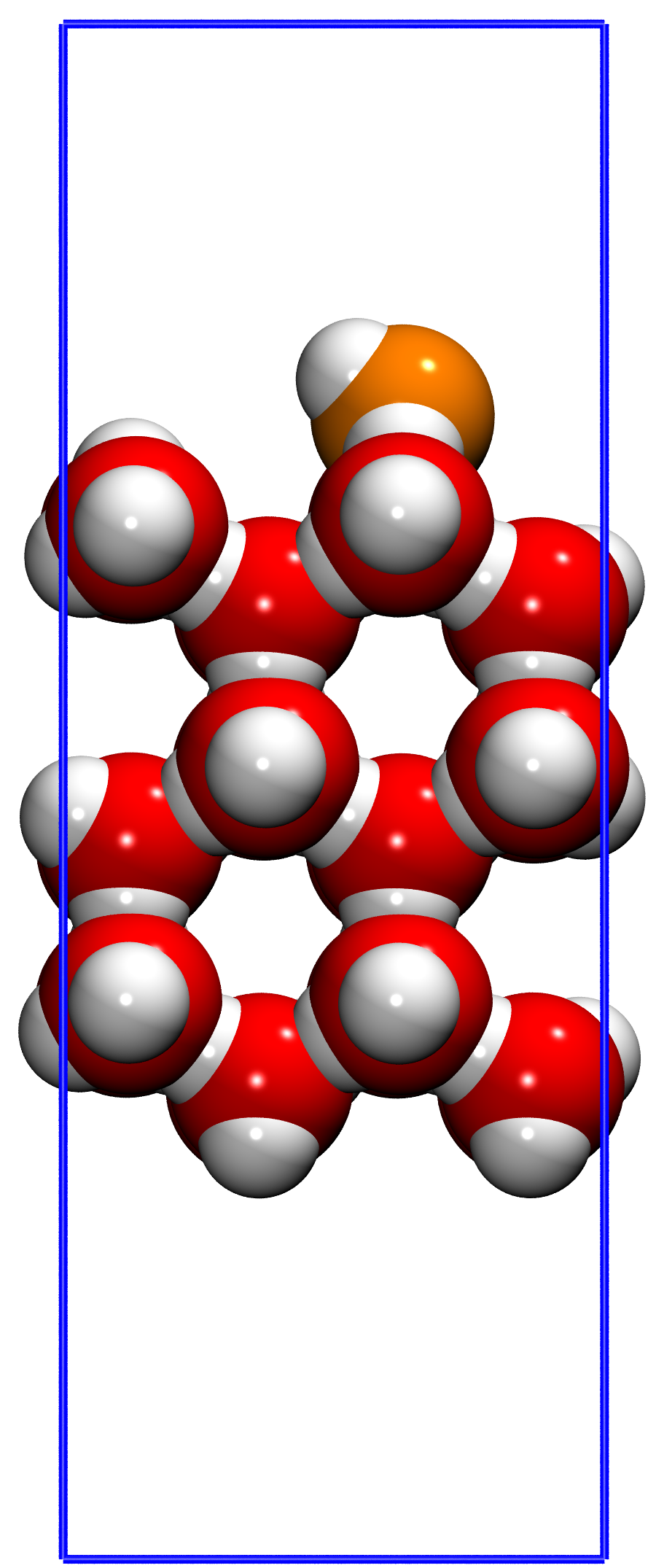}
}
\hfill
\subfigure[Water Cry Des.\label{fig:cry_samples_des}]{
\includegraphics[width=0.15\columnwidth]{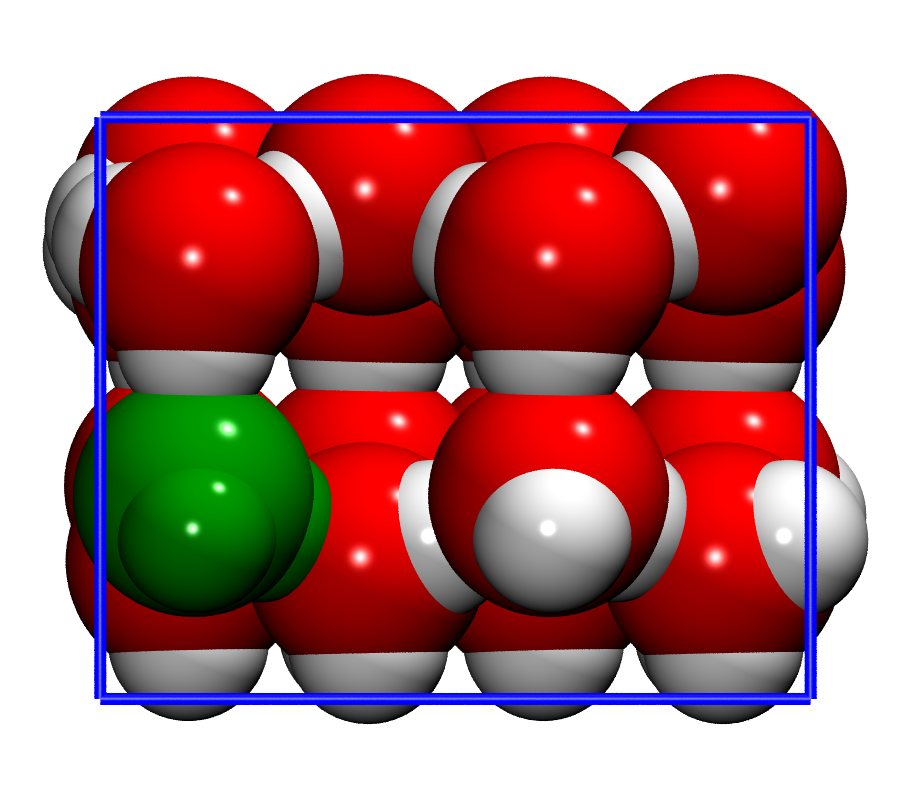}
\includegraphics[width=0.15\columnwidth]{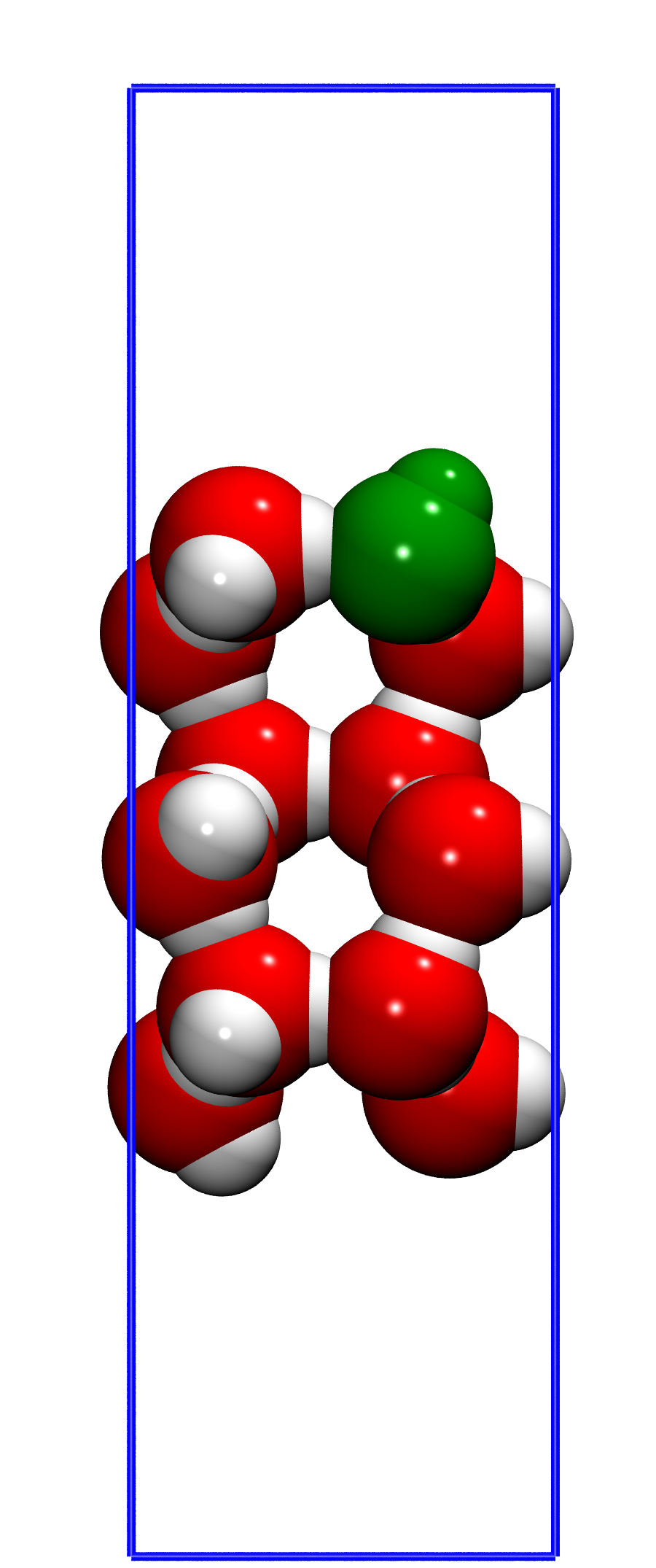}
\includegraphics[width=0.15\columnwidth]{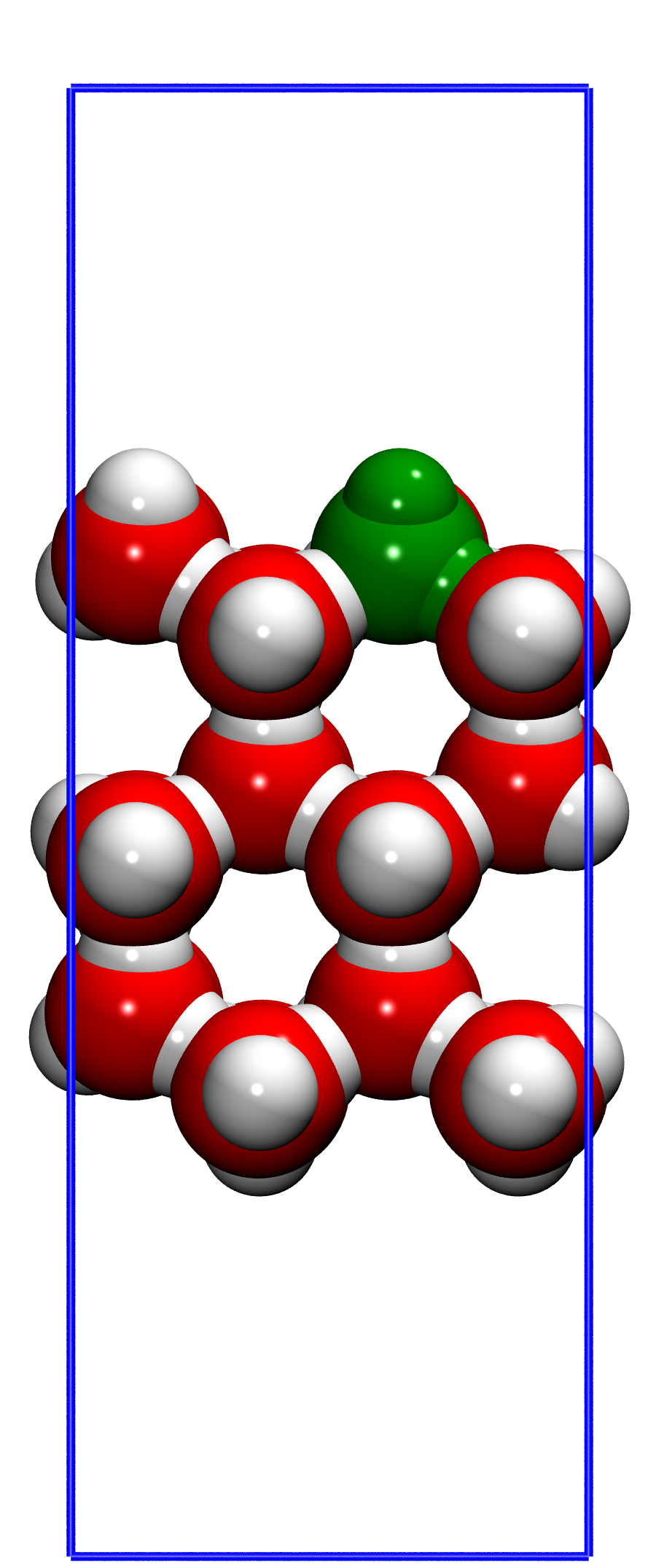}
}
\hfill
\subfigure[Ammonia Cry Ads.]{
\includegraphics[width=0.15\columnwidth]{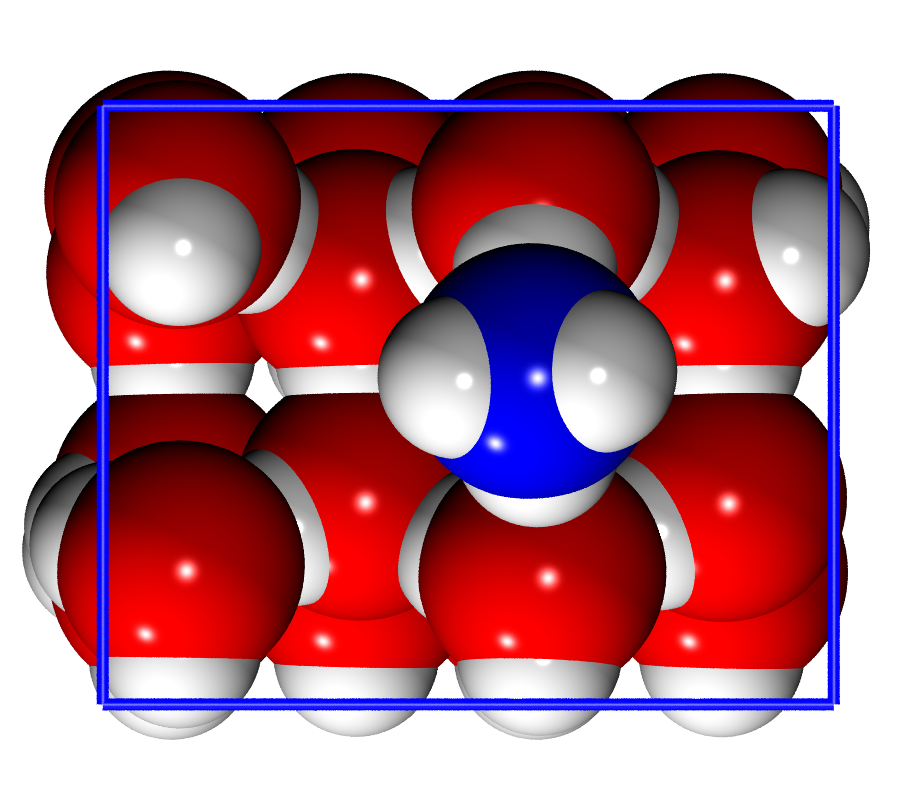}
\includegraphics[width=0.145\columnwidth]{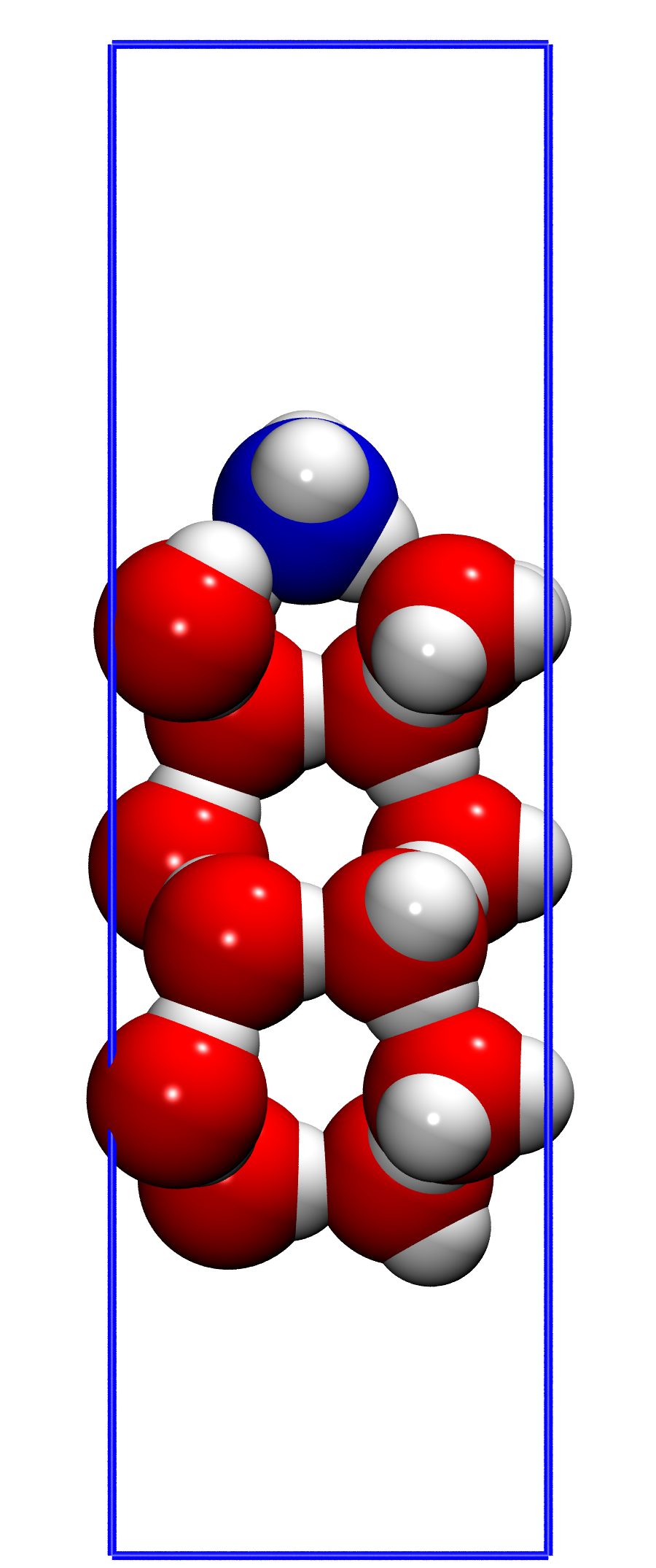}
\includegraphics[width=0.145\columnwidth]{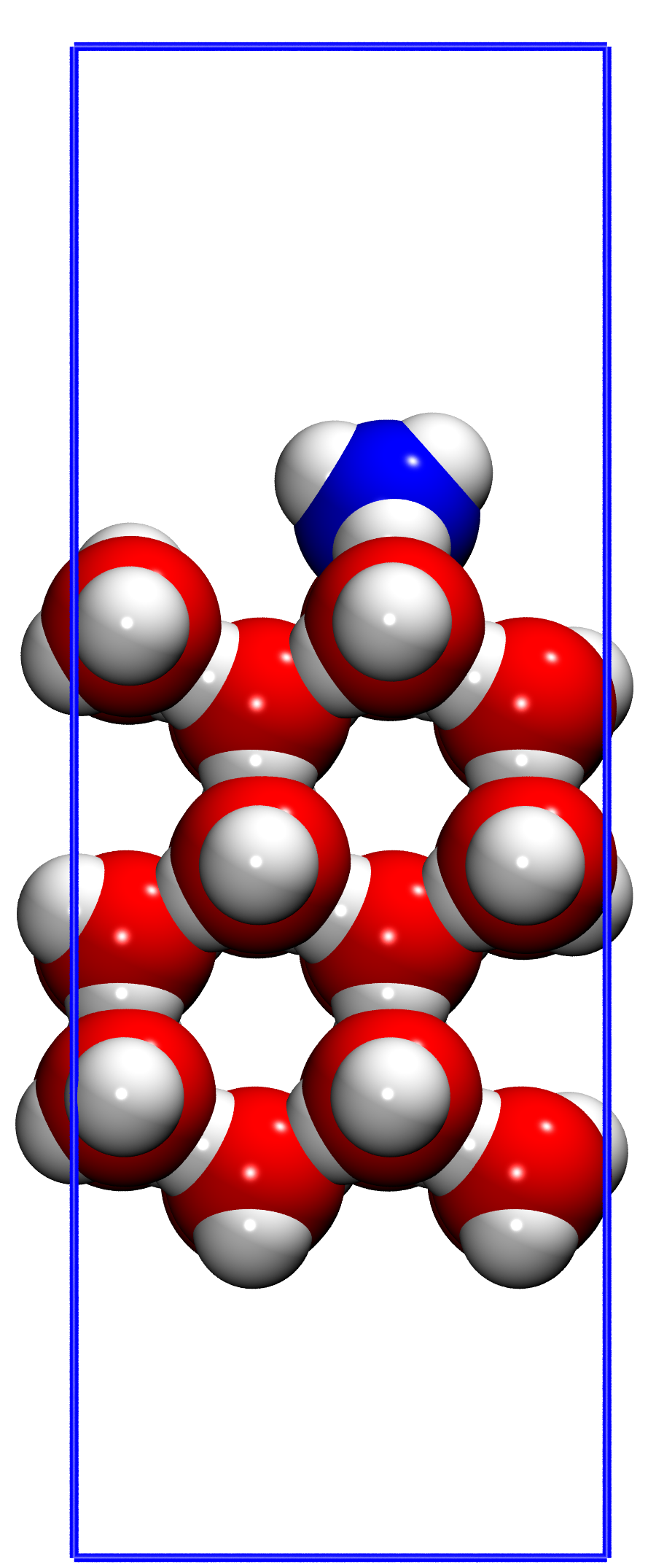}
}
\hfill
\subfigure[Methanol Cry Ads.]{
\includegraphics[height=0.14\columnwidth]{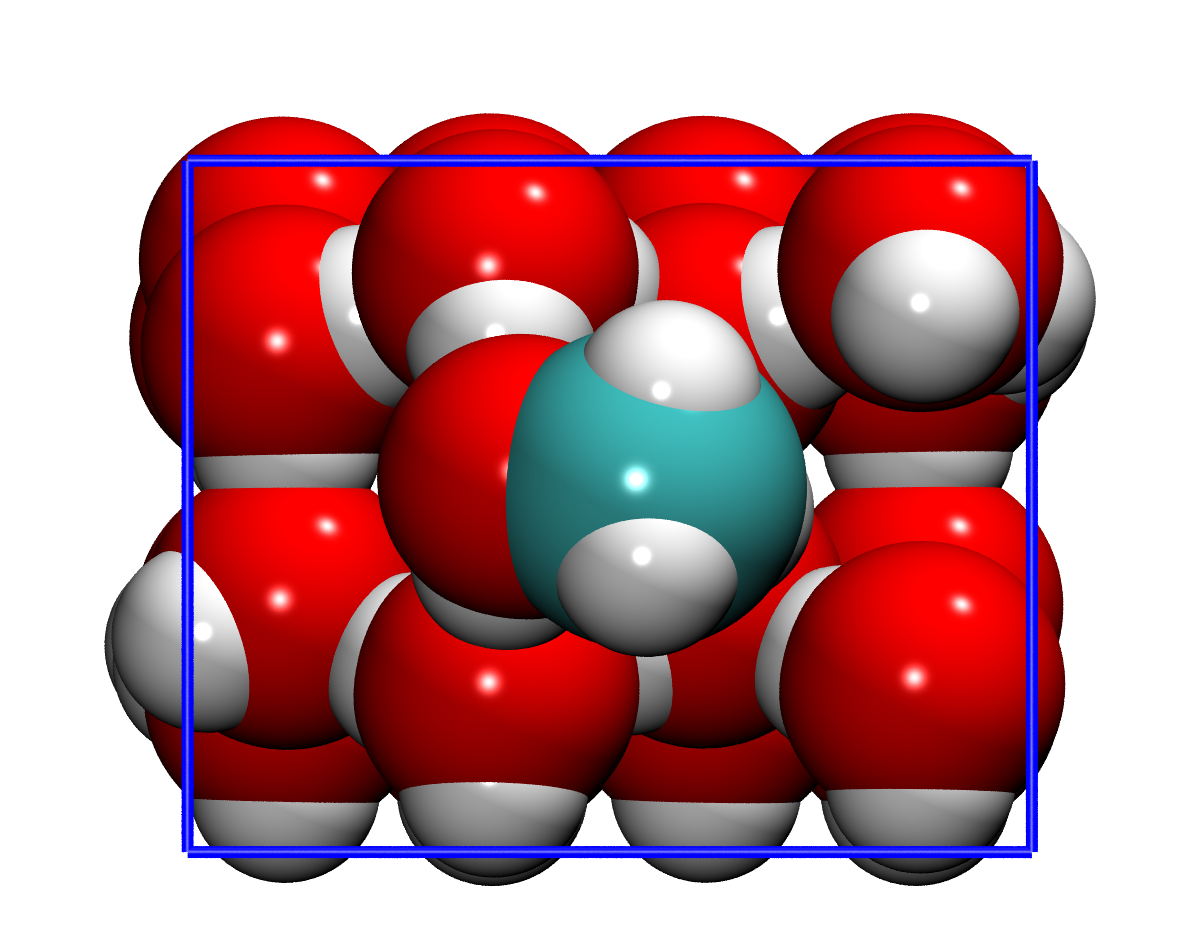}
\includegraphics[height=0.352\columnwidth]{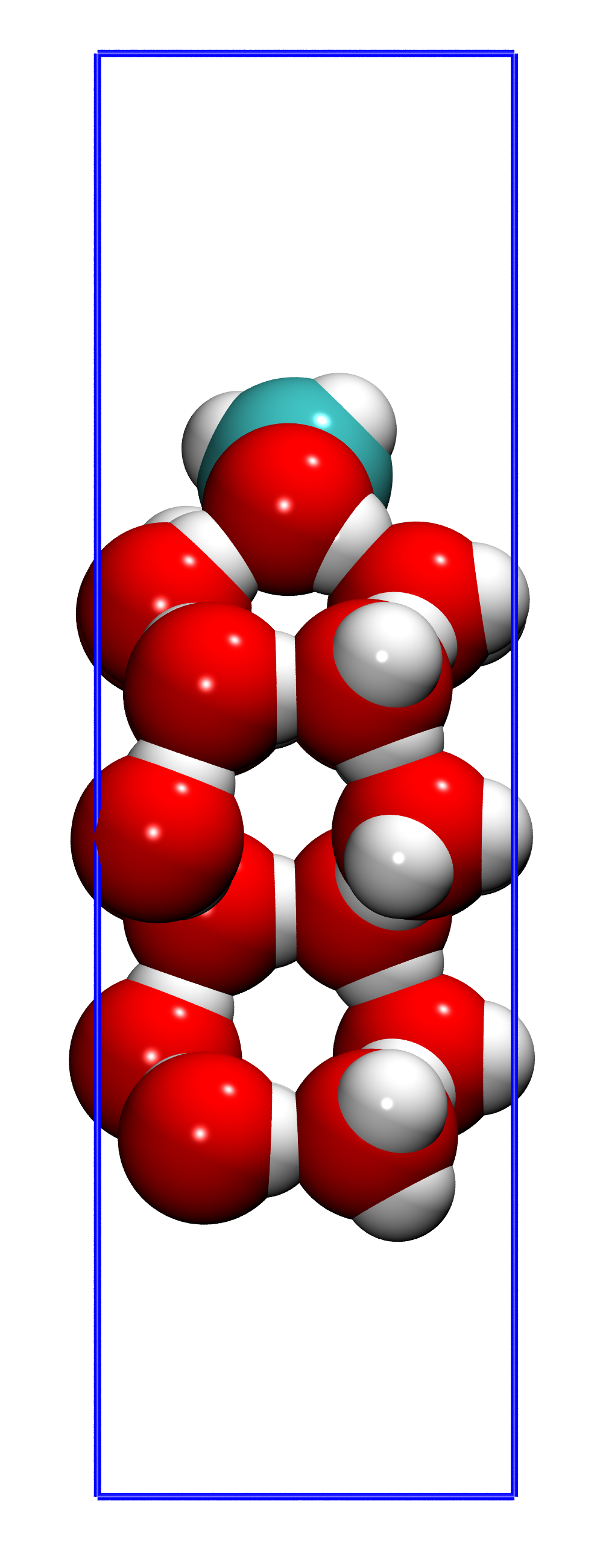}
\includegraphics[height=0.352\columnwidth]{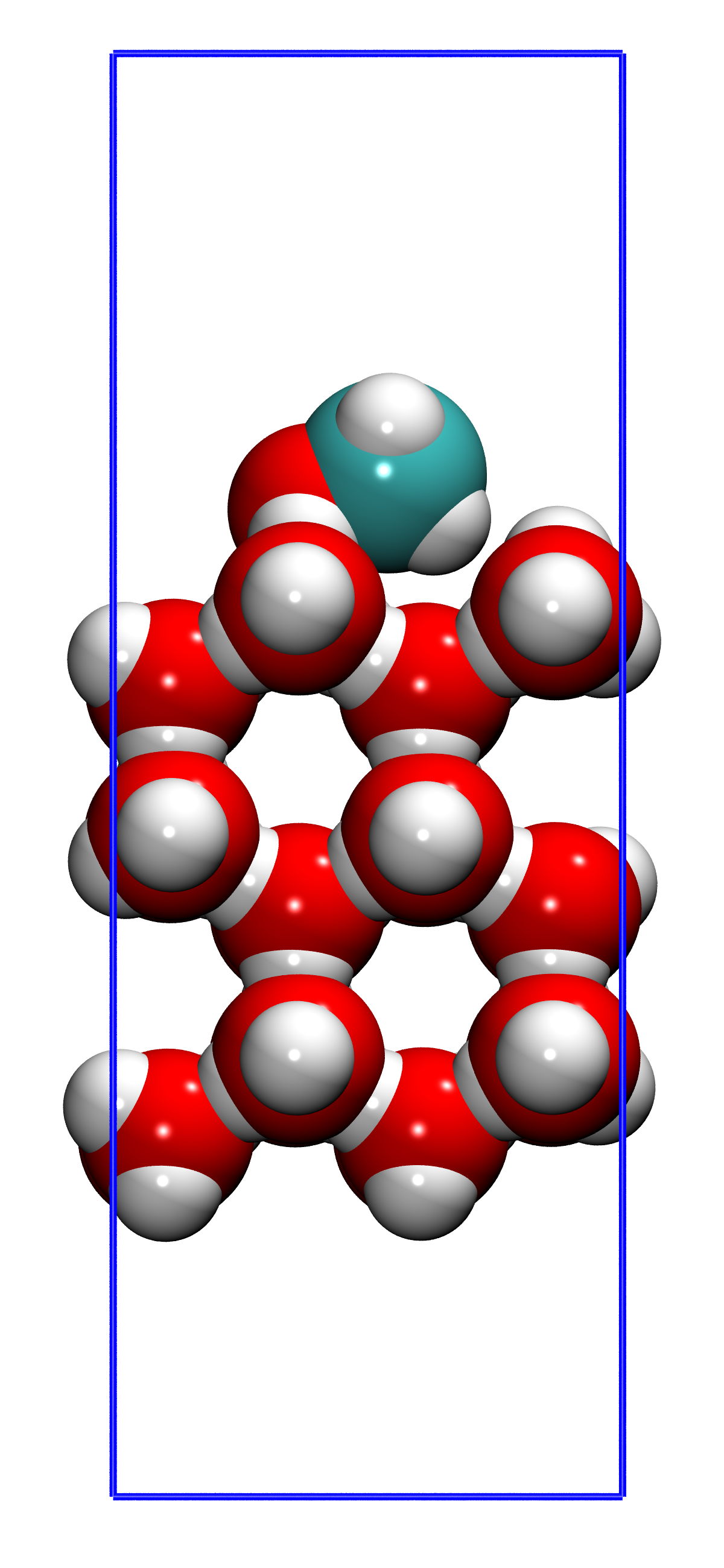}
}
\end{center}
\caption{Optimized geometries of the crystalline (Cry) ice samples with three different orthogonal perspectives of the periodic cell: water (a, b), ammonia (c) and methanol (d). 
Atom color legend: oxygen in red, hydrogen in white, nitrogen in blue and carbon in cyan.}
\label{fig:cry_samples}
\end{figure*}

The amorphous ice grain models for molecular calculations are those containing 200 water molecules already used in Refs.~\onlinecite{Aurele_grain,Tinacci_NH3_BE}, as well as large part of the methodology, in particular for what regards the ONIOM approach.\cite{Tinacci_NH3_BE} 
Small modifications to the above references are the computational code and the DFT method used to carry out geometry optimizations and frequency calculations. 
Specifically, we relied on the ORCA program (v.5.0.2),\cite{ORCA} which implements a full integration with the xTB code,\cite{xtb,GFN2} used for the treatment of the low-level zone. The ONIOM boundary has been set to 5 {\AA} with respect to the center of mass of the adsorbate, as well as the fixing of the atomic positions. In other words, the high-level method coincides to the atoms free to move, while the low-level method to the atomic coordinates fixed during the optimization procedure. As regards the DFT high-level method, the B97-3c \cite{b97_3c} functional was used, which already incorporates a very well-balanced basis set, without the need for any BSSE correction. The ONIOM energies were refined using the DLPNO-CCSD(T) \cite{DLPNO_CCSD(T)_new} coupled with the aug-cc-pVTZ \cite{kendall1992dunning} as primary basis set and the aug-cc-pVTZ/C\cite{weigend2002efficient} as auxiliary basis set for the resolution of identity (RI) approximation in electron repulsion integrals. "Tight PNO" and "Tight SCF" settings were used in DLPNO-CCSD(T) calculations. All the DLPNO-CCSD(T) calculations were corrected for the BSSE using the counterpoise method.\cite{Counterpoise} As described in our previously published paper,\cite{Tinacci_NH3_BE, tinacci_2023_water, bariosco2024binding, bariosco2025methanol} every sample has its own reference isolated surface obtained by deleting the adsorbed molecule in the complex and then re-optimizing it. Frequency calculations were calculated by numerical differentiation of the first analytical derivatives using the center-difference formula (\textit{i.e.} two displacements for each atom in each direction).

\paragraph*{Computed cases:}
Three adsorption sites were studied for three different molecules: water, ammonia and methanol. 
In Table~\ref{tab:be_mol} and in Fig.~\ref{fig:mol_samples} the energetic and structural features of these samples are presented.

\begin{table}
\caption{\label{tab:be_mol} Binding enthalpies of molecular (Mol) cases and their decomposition.
All values are in kJ/mol.}
\begin{ruledtabular}
\begin{tabular}{lccccc}
Species & Sample & BH(0) & BE$_e$ & $\Delta$ZPE & $\delta$E$_{def}$ \\
\hline
\multirow{3}{*}{H$_2$O}   & Mol 1                   & 30.9 & 45.3 & -10.7 &  3.7 \\
                          & Mol 2                   & 44.6 & 71.7 & -12.0 & 15.1 \\
                          & Mol 3                   & 56.6 & 84.6 & -15.1 & 12.9 \\
\hline
\multirow{3}{*}{NH$_3$}   & Mol 1                   & 24.3 & 57.8 &  -5.8 & 27.8 \\
                          & Mol 2                   & 34.1 & 54.1 &  -9.1 & 10.9 \\
                          & Mol 3                   & 44.5 & 69.1 & -10.8 & 13.9 \\
\hline
\multirow{3}{*}{CH$_3$OH} & Mol 1                   & 20.3 & 31.8 & -10.2 &  1.3 \\
                          & Mol 2                   & 35.9 & 54.2 &  -8.4 &  9.9 \\
                          & Mol 3                   & 55.1 & 75.3 &  -9.6 & 10.6 \\
\end{tabular}
\end{ruledtabular}
\end{table}

\subsection{Periodic calculations}\label{app:comp_periodic}

\begin{table}
\caption{\label{tab:be_cry} Binding enthalpies of crystalline (Cry) samples and their decomposition.
All values are in kJ/mol.}
\begin{ruledtabular}
\begin{tabular}{lccccc}
Species & Sample & BH(0) & BE$_e$ & $\Delta$ZPE & $\delta$E$_{def}$ \\
\hline
\multirow{2}{*}{H$_2$O}   & Cry Ads & 50.4 & 63.7 & -13.4 &  5.2 \\
                          & Cry Des & 71.1 & 86.5 & -15.3 & 15.7 \\
\hline
NH$_3$                    & Cry Ads & 55.2 & 67.5 & -12.3 &  7.5 \\
\hline
CH$_3$OH                  & Cry Ads & 60.1 & 70.2 & -10.1 &  6.0 \\
\end{tabular}
\end{ruledtabular}
\end{table} 

\begin{center}
\begin{figure}
\subfigure[Fragment for "Cry Ads" sample.\label{fig:cry_frag_ads}]{
\includegraphics[height=0.135\columnwidth]{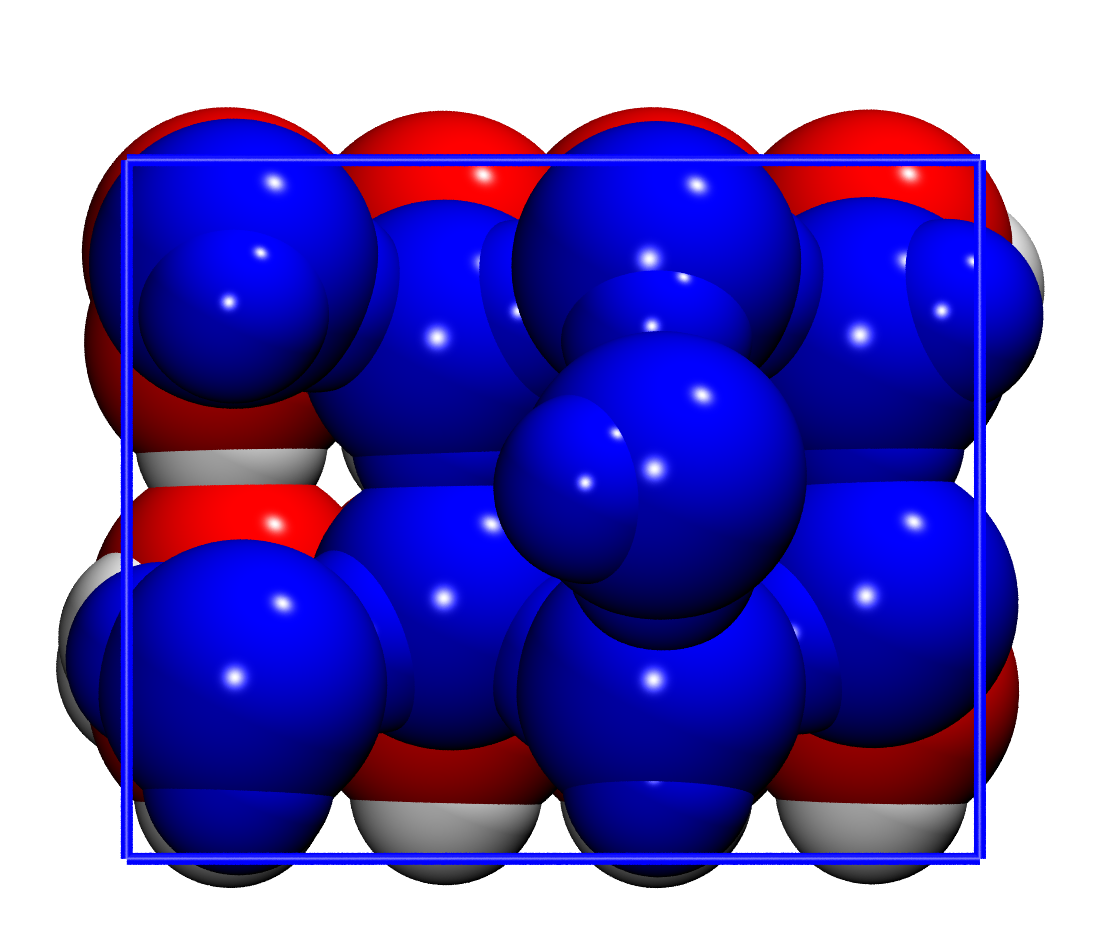}
\includegraphics[height=0.335\columnwidth]{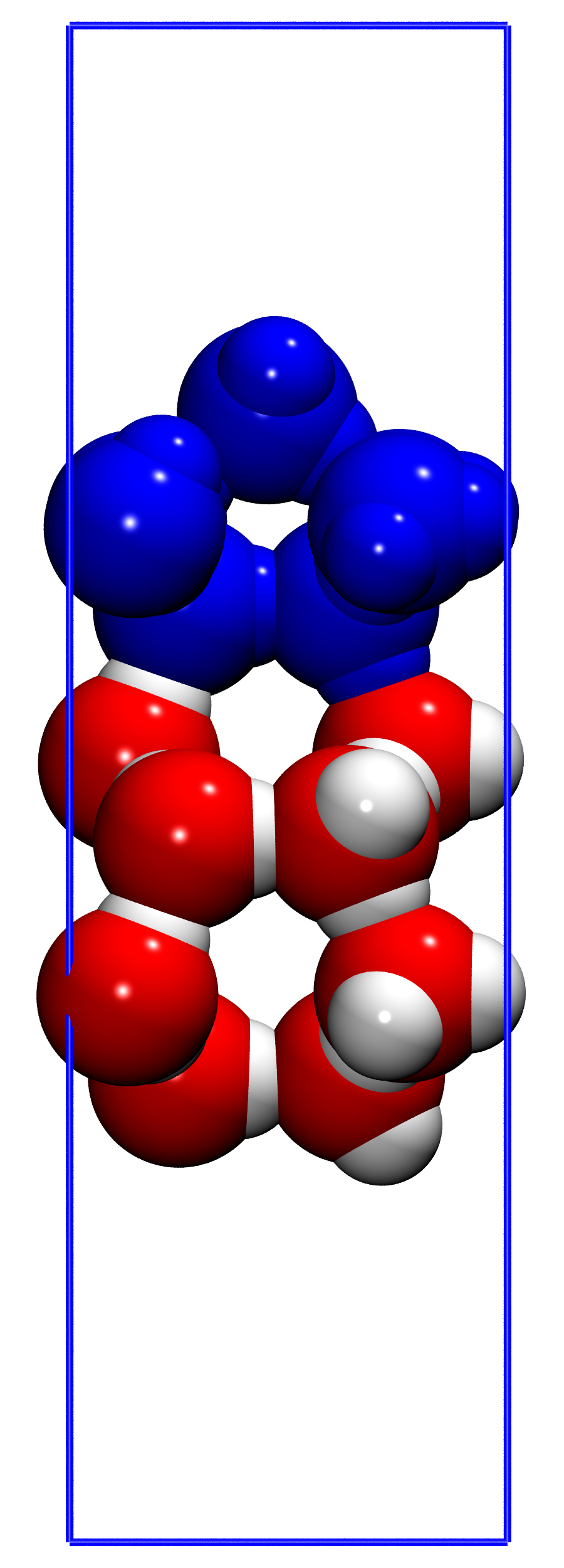}
\includegraphics[height=0.335\columnwidth]{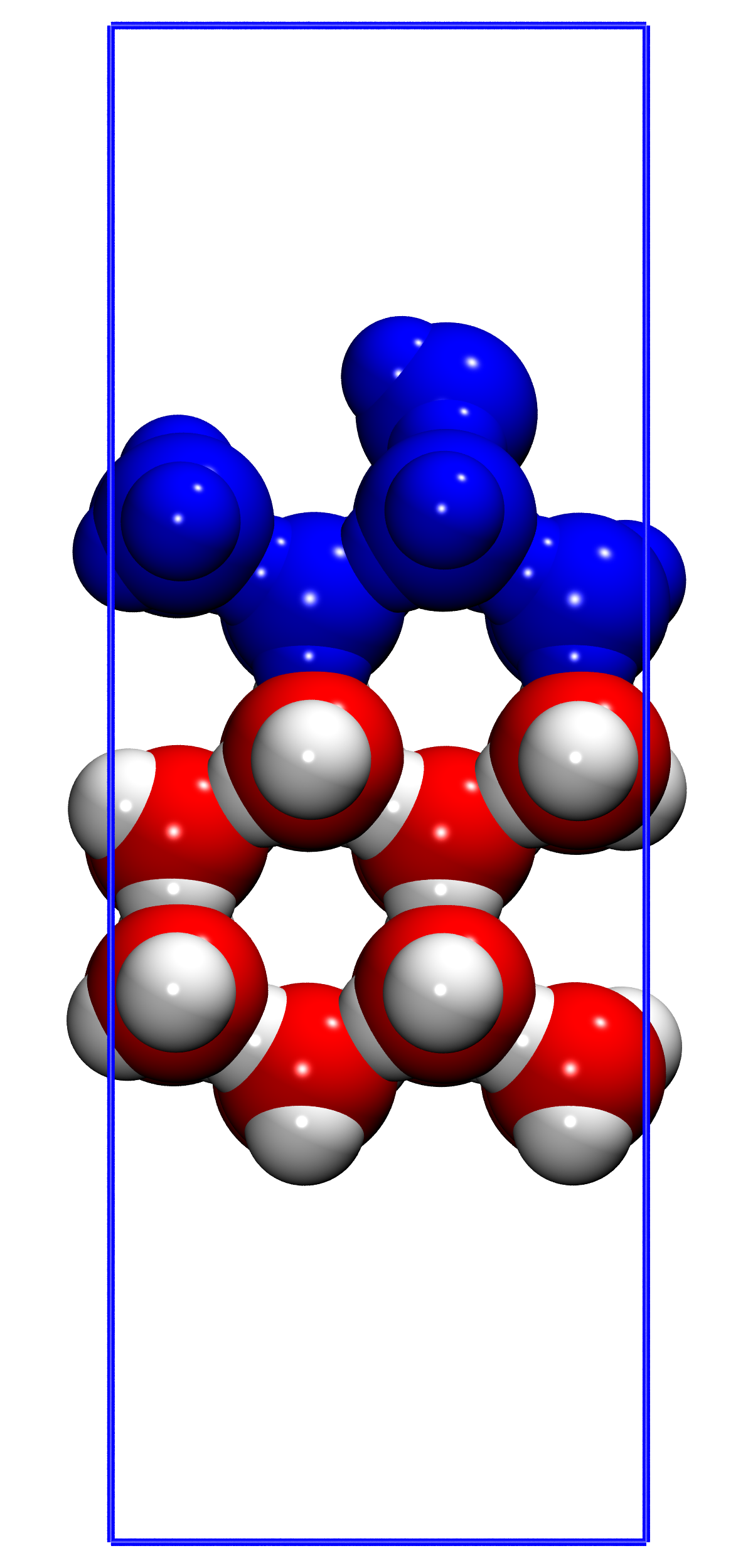}
}
\hfill
\subfigure[Fragment for "H$_2$O Cry Des" sample.\label{fig:cry_frag_des}]{
\includegraphics[height=0.135\columnwidth]{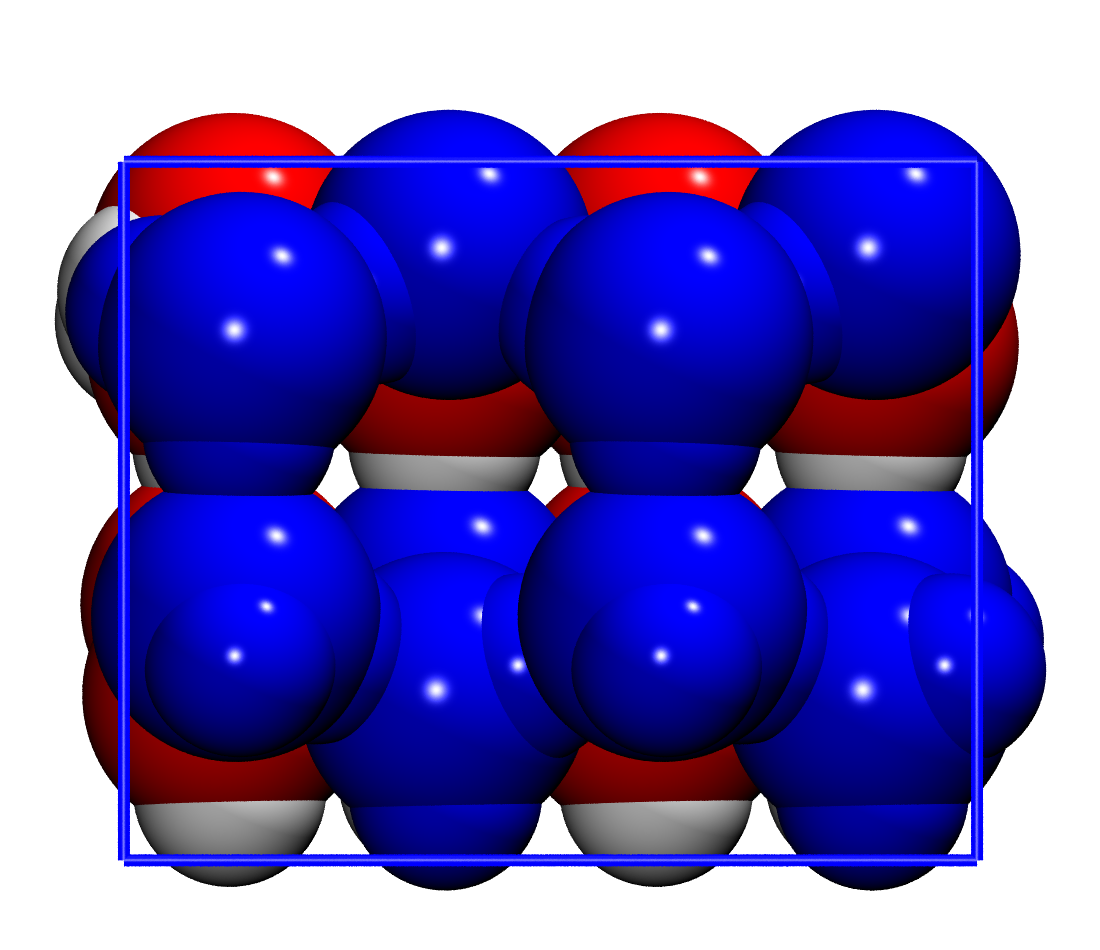}
\includegraphics[height=0.335\columnwidth]{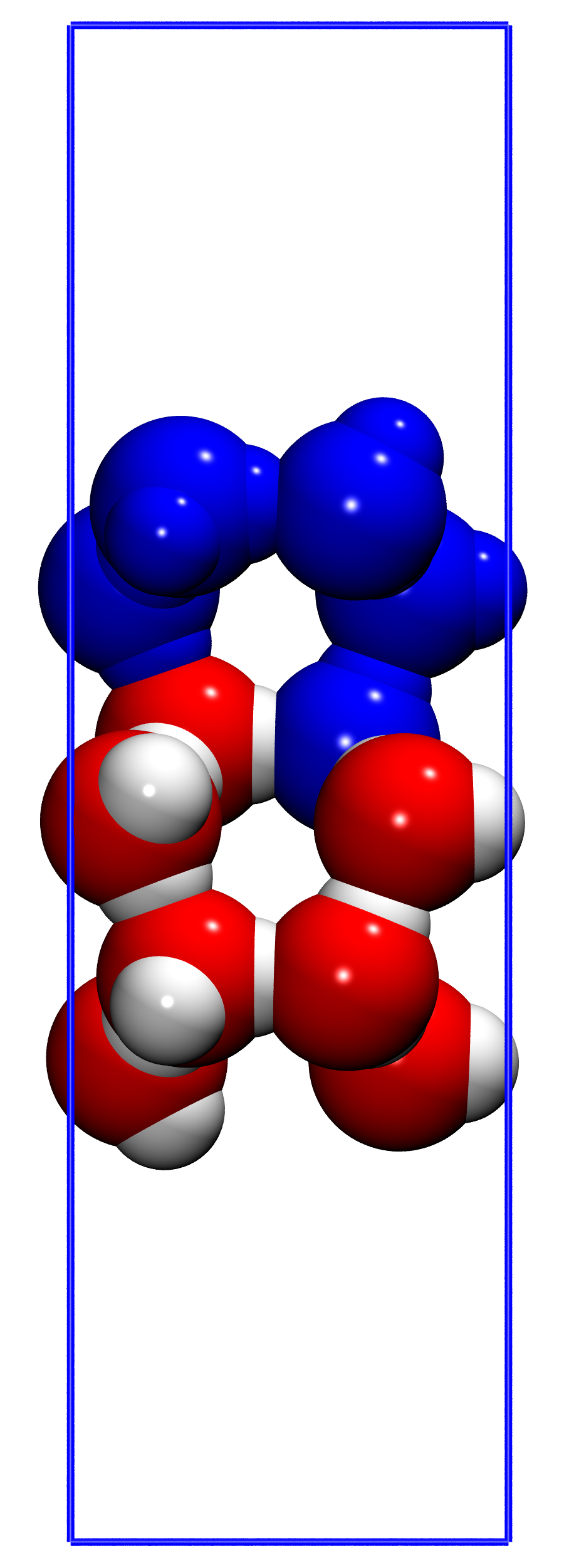}
\includegraphics[height=0.335\columnwidth]{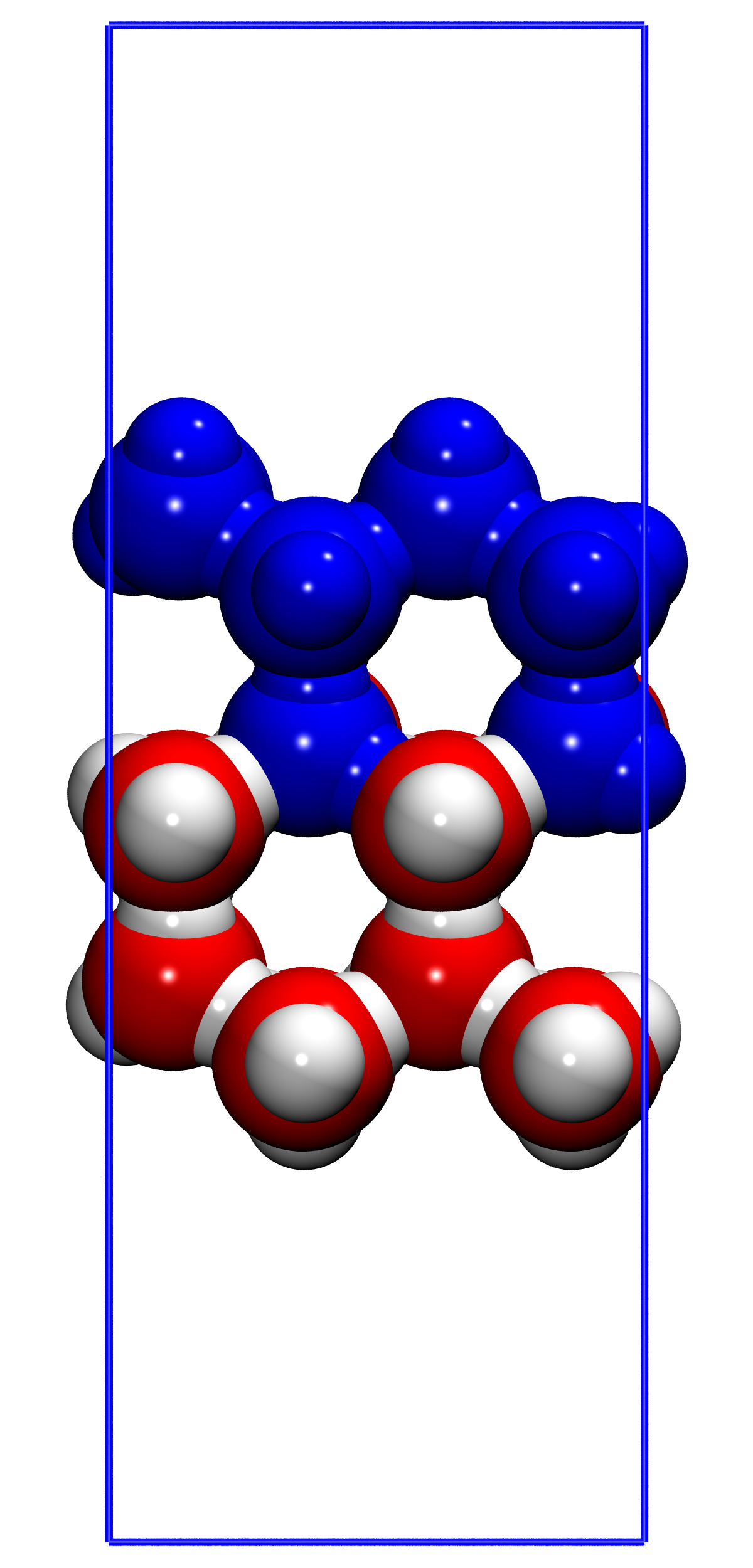}
}
\caption{Orthogonal perspectives of the periodic cell of "Cry Ads" (a) (for the water case) and "Cry Des" (b) samples. In blue the water molecules free to move in the "Fragment" frequency calculations. Atom color legend: oxygen in red, hydrogen in white.}
\label{fig:cry_fragment}
\end{figure}
\end{center}

Periodic calculations adopting the Ih proton ordered ice and its surface models were carried out with the CRYSTAL17 code.\cite{crystal17,crystal} 
To be consistent with molecular calculations the B97-3c method was used.\cite{b97_3c} 
The k-point grid was generated via the Monkhorst-Pack scheme, and the shrinking factor was set to 6, for a total number of 21 independent k-points in the irreducible First Brillouin Zone (FBZ). 
Tolerances of the integral calculation were decreased to 10$^{-7}$ (10$^{-6}$ default) for Coulomb overlap and Coulomb penetration in the direct space. 
The DFT integration grid was set to 99 radial points and 1454 angular points (XXLGRID) in a Lebedev scheme. 
The SCF tolerance on energy convergence was increased to 10$^{-11}$ Hartree. 
During bulk geometry optimizations, both atomic positions and cell vectors were relaxed, setting the tolerances for the convergence of the maximum allowed gradient and the maximum atomic displacement to $4 \cdot 10^{-5}$ Hartree/Bohr and $12 \cdot 10^{-5}$ Bohr, respectively.
Frequency calculations at the $\Gamma$-point have been performed with the central-difference formula (\textit{i.e.} two displacements for each atom in each cartesian direction).

In order to obtain a more reliable vibrational correction of the cohesive energy, phonon dispersion of the bulk structure was evaluated by means of the supercell (2x2x2) approach (the shrinking factor was accordingly reduced to 3), thus breaking the translational symmetry and allowing for a larger sampling of the phonon modes than in $\Gamma$-point only. 

To simulate the ice surface, the bulk structure was cut along the [100] direction, ranging the thickness from 1 up to 4 layers, in order to identify the minimum thickness that leads to converged surface properties (see Supplementary Material). The atomic coordinates of the surface were set free to optimize, while cell parameters have been fixed to the optimized bulk values.

Harmonic frequency calculations at the $\Gamma$-point have been performed for all the atoms ("Full") while the “Fragment” model only includes the first coordination shell of the water molecules closer to the adsorbate (see Fig. \ref{fig:cry_fragment}). The aim is to compare the results obtained from a full frequency calculation versus a smaller fragment, in terms of the ZPE correction and the resulting partition functions, in order to save large part of the computational cost required by a full frequency calculation. It is important to realize that the fragment calculations include the energetic contribution of all atoms of the whole system when computing the nuclei displacements of the fragments and, therefore, the influence of the surrounding is taken into account. Only the direct coupling between vibrational modes of the fragment with the remaining ones of the system is ignored.

\paragraph*{Computed cases:}
The molecules are the same as those used for molecular calculations, however, on the crystalline surface in the case of water, two different desorption processes were simulated: the first one by removing the adsorbed water molecule on the (100) surface ("Cry Ads", see Fig.~\ref{fig:cry_samples}a), the second one by removing a molecule from the bare (100) ice surface as cut out from the bulk ("Cry Des", see Fig.~\ref{fig:cry_samples}b, the removed water molecule is highlighted in  green). 
In Table~\ref{tab:be_cry} and in Fig.~\ref{fig:cry_samples} the energetic and structural features of these samples are presented.

Rendering of all the molecule images have been obtained via the VMD software.\cite{vmd}

\section{Vibrational-Rotational partition function for low frequency modes}\label{app:q_vib_appendix}

\begin{figure*}[!ht]
\begin{center}
\subfigure[Water]{
\includegraphics[width=0.99\textwidth]{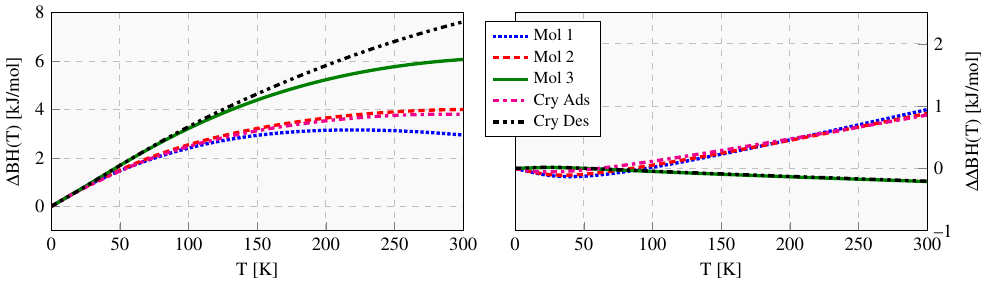}
}
\subfigure[Ammonia]{
\includegraphics[width=0.99\textwidth]{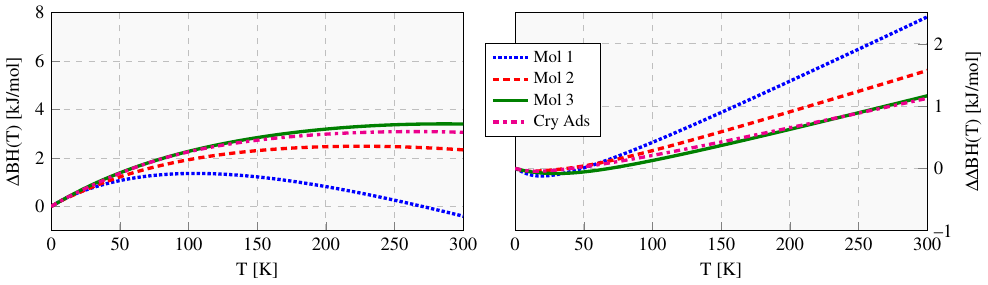}
}
\subfigure[Methanol]{
\includegraphics[width=0.99\textwidth]{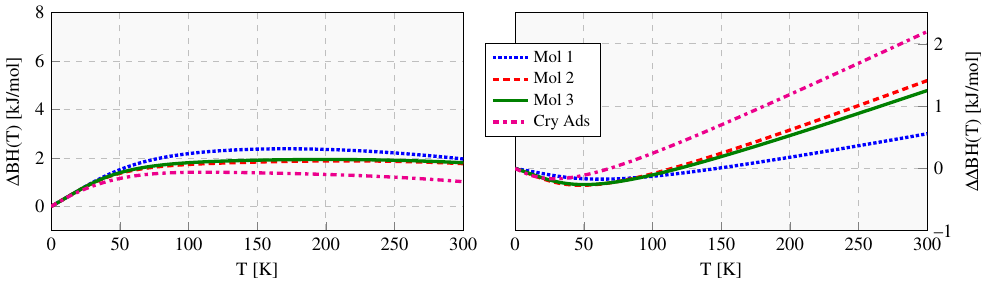}
}
\end{center}
\caption{\textit{Left panels:} Thermal correction of BE ($\Delta$BH(T) = BH(T) - BH(0)) as a function of the temperature (Eq.~\ref{eq:BE_thermal}) for all the cases of Table \ref{tab:be}: water (a), ammonia (b), methanol (c).
\textit{Right panels:} Vibrational enthalpy difference between the RRHO (Eq.~\ref{eq:rrho}) and quasi-RRHO (Eq.~\ref{eq:quasi_rrho}) approaches, \textit{i.e.} $\Delta\Delta$BH(T) = $\Delta$BH(T)(RRHO) - $\Delta$BH(T)(q-RRHO).}
\label{fig:bh_t}
\end{figure*}

The results of the thermal correction as a function of temperature for all the cases presented in Table \ref{tab:be_mol} and \ref{tab:be_cry} are shown in Fig.~\ref{fig:bh_t}.

As already said in the main text, thermal corrections at high temperatures (close to the desorption, \textit{i.e.} $\approx$ 200 K) are not negligible (up to 6 kJ/mol), depending on the temperature. 

In Fig.~\ref{fig:bh_t}, right panels, we also studied the differences between full RRHO (Eq.~\ref{eq:rrho}) and q-RRHO (Eq.~\ref{eq:quasi_rrho}) corrections to the BH(0). As one can see, the $\Delta\Delta$BH(T) correction, \textit{i.e} $\Delta$BH(T)(RRHO) - $\Delta$BH(T)(q-RRHO) is very low at all studied temperatures.

Moreover, using the same correction for low-frequency vibrations of Eq.~\ref{eq:quasi_rrho} and the correction for the entropy proposed by Grimme et al.,\cite{pracht2021calculation} we recalculate the vibrational partition function:
\begin{equation}\label{eq:q_vib_q_rot}
    m-q_{vib} = \Pi_i^{3N-6} \omega(\nu_i)~ q_{vib}(\nu_i) + (1- \omega(\nu_i))~ q_{rot}(\nu_i) \; .
\end{equation}
where $\omega(\nu_i)$ is the dumping function defined in Eq.~\ref{eq:omega} and $q_{rot}(\nu_i)$ is the $\nu_i$ normal mode weighted rotational partition function, \textit{i.e.}:
\begin{equation}
    q_{rot}(\nu_i) = \frac{\sqrt{8 ~\pi^2~ k_B ~T}}{h} \sqrt{\frac{\mu(\nu_i) ~I_{av}}{\mu(\nu_i) + I_{av}}} \; ,
\end{equation}
$I_{av}$ is the average inertia tensor and $\mu(\nu_i)$ is equal to $\frac{h}{8 \pi \nu_i}$.

\begin{figure}
\begin{center}
\includegraphics[width=0.99\columnwidth]{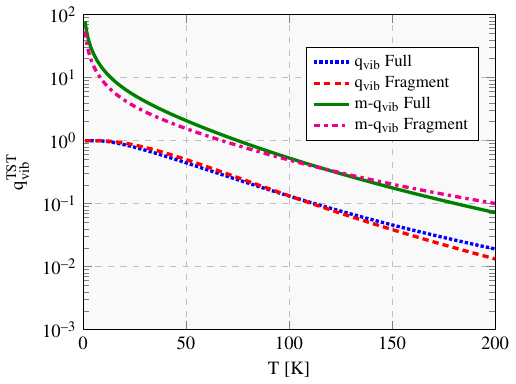}
\end{center}
\caption{Vibrational partition function ratio (Eq.~\ref{eq:q_vib_ratio}) for methanol in the crystalline (Cry Ads) case. "Full" and "Fragment" stand respectively for complete, and desorbing molecule only vibrational frequencies calculations. q$_{vib}$ and m$-$q$_{vib}$ are calculated according to Eqs. \ref{eq:q_vib} and \ref{eq:q_vib_q_rot}.}
\label{fig:q_vib_rot}
\end{figure}

In Fig.~\ref{fig:q_vib_rot} the $q_{vib}^{TST}$ ratios for desorption mechanism (Eq.~\ref{eq:q_vib_ratio}) are shown using the two different ways to compute the vibrational partition function, \textit{i.e.} the standard approach (Eq.~\ref{eq:q_vib}) and the corrected low motion modes for rotations (Eq.~\ref{eq:q_vib_q_rot}, m$-$q$_{vib}$ in Fig.~\ref{fig:q_vib_rot}). From the plot it is clear that the $q_{vib}^{TST}$ calculated using the m$-$q$_{vib}$ approach fails in reproducing the behaviour for T$\rightarrow$0, diverging from q$_{vib}$.

Due to above-mentioned results, \textit{i.e.} the small differences in the thermal correction and the divergence at low temperatures, we suggest, if needed, to use the normal RRHO to correct the BH(0).

\section{Kooij-Arrehnius fitting}\label{app:kooij}

\begin{figure}
\begin{center}
\subfigure[Desorption rate ratio.]{
\includegraphics[width=\columnwidth]{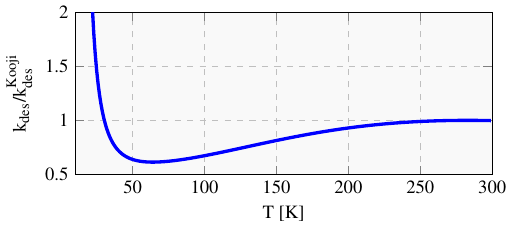}
}
\subfigure[Numerical TPD spectra.]{
\includegraphics[width=\columnwidth]{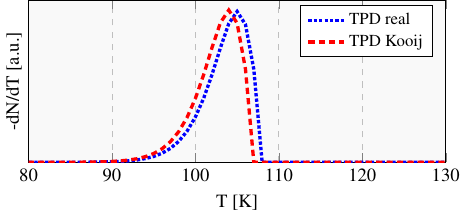}
}
\end{center}
\caption{(a) Desorption rates ratio between the full TST approach ($k_{des}(T)$, Eq. \ref{eq:k_des}) and the fitted one by the Kooij-Arrehnius equation ($k_{des}^{Kooij}(T)$, Eq. \ref{eq:kooij}) as a function of the temperature. (b) Numerical TPD spectra computed as first-order desorption equation. The considered case is the water Mol 1 sample (Table \ref{tab:be}).}
\label{fig:kooij_fit}
\end{figure}

In this section we show the accuracy of the Kooij fitting (Eq.~\ref{eq:kooij}) to the desorption rate, taking as test case the TPD spectrum of water Mol 1 sample. This allows to easily derive $\nu(T)$ and BH(T) a posteriori from the fitted TPD spectrum. 
Fig.~\ref{fig:kooij_fit}a shows the ratio between the Kooij-Arrehnius fitting and the real one: the ratio is close to 1 above 25 K, while at lower temperatures it tends to diverge. This means that in the range of the desorption temperatures the fitting procedure is accurate. 
Indeed, the overlap of the two numerical TPD spectra of Fig.~\ref{fig:kooij_fit}b is very good (taking the same conditions as those used for the TPD spectra of Fig.~\ref{fig:TPD_cluster}); the difference in the resulting T$_{peak}$ between the two approaches is less then 1 K.
The fitting parameters, compared with the pre-exponential factor and binding energy from Table \ref{tab:be}, are: $\alpha = 1.3\cdot 10^{15}$ $s^{-1}$ ($\nu_{TST}^{vib}(T_{peak}) = 5.9 \cdot 10^{14}$ $s^{-1}$), and $\gamma = 31.7$ kJ/mol (BH(T) = 33.8 kJ/mol).

\clearpage

\end{document}